\documentclass[%
 reprint,
superscriptaddress,
nofootinbib,
 amsmath,amssymb,
 aps,
 prd,
]{revtex4-1}
\usepackage{bbold}
\usepackage[breaklinks=true,colorlinks=true
,urlcolor=blue
,anchorcolor=blue
,citecolor=blue
,filecolor=blue
,linkcolor=blue
,menucolor=blue
,pagecolor=blue
,linktocpage=true
,pdfproducer=medialab
,pdfa=true
]{hyperref}
\usepackage{graphicx,hyperref,color}
\usepackage{color}
\usepackage[dvipsnames]{xcolor}
\usepackage{graphicx}
\usepackage{dcolumn}
\usepackage{bm}
\usepackage{bbm}
\usepackage{mathtools}
\usepackage{amsmath,amssymb}
\usepackage{physics}
\usepackage{float}

\usepackage{tikz}
\usetikzlibrary{arrows,arrows.meta,intersections, calc,positioning,decorations.pathreplacing,decorations.pathmorphing,shapes}
\usetikzlibrary{patterns}
\usetikzlibrary{decorations.markings}
\usetikzlibrary{knots}

\def\n{{\nu}}
\def\ep{{\epsilon}}

\def\frac#1#2{{#1\over #2}}

\def\s{\sqrt}

\def\be{\begin{equation}}
\def\ee{\end{equation}}
\def\ba{\begin{eqnarray}}
\def\ea{\end{eqnarray}}

\def\no{\nonumber \\}

\def\ep{\epsilon}

\usepackage{braket}

\newcommand{\tq}{{\mathtt q}}

\newcommand{\tp}{{\mathtt p}}
\def\CE{{\mathcal E}}

\allowdisplaybreaks[4]

\begin{document}
\begin{flushright}
YITP-25-87
\\
\end{flushright}
\title{{\large Multi-entropy and the Dihedral Measures at Quantum Critical Points}}

\author{Jonathan Harper}
\affiliation{Center for Gravitational Physics and Quantum Information, Yukawa Institute for Theoretical Physics, Kyoto University, Kitashirakawa Oiwakecho, Sakyo-ku, Kyoto 606-8502, Japan}

\author{Ali Mollabashi}
\affiliation{ School of Quantum Physics and Matter, Institute for Research in Fundamental Sciences (IPM), 19538-33511, Tehran, Iran}

\author{Tadashi Takayanagi}
\affiliation{Center for Gravitational Physics and Quantum Information, Yukawa Institute for Theoretical Physics, Kyoto University, Kitashirakawa Oiwakecho, Sakyo-ku, Kyoto 606-8502, Japan}
\affiliation{Inamori Research Institute for Science, 620 Suiginya-cho, Shimogyo-ku, Kyoto 600-8411, Japan}

\author{Kenya Tasuki}
\affiliation{Center for Gravitational Physics and Quantum Information, Yukawa Institute for Theoretical Physics, Kyoto University, Kitashirakawa Oiwakecho, Sakyo-ku, Kyoto 606-8502, Japan}


\begin{abstract}
The multi-entropy and dihedral measures are a class of tractable measures for multi-partite entanglement, which are labeled by the Rényi index (or replica number) $n$ as in the Rényi entanglement entropy. The purpose of this article is to demonstrate that these quantities are new useful probes of quantum critical points by examining concrete examples. 
In particular, we compute the multi-entropy and dihedral measures in the $1+1$ dimensional massless free scalar field theory on a lattice and in the transverse-field Ising model. For $n=2$, we find that the numerical results in both lattice theories quantitatively agree with those from conformal field theoretic calculations. For $n=3$ and $n=4$, we provide new predictions of these measures for the massless scalar field theory.

\end{abstract}

\maketitle




\section{Introduction}

The entanglement entropy has played a major role as it provides a universal characterization of physical properties of quantum many-body systems in various subjects such as high-energy physics, quantum gravity, and condensed matter physics. The entanglement entropy is defined for any bi-partite system and has the advantage that it is uniquely defined as the von-Neumann entropy and that it can be computed straightforwardly in quantum many-body systems. It plays a role of a quantum order parameter to detect the quantum phase transitions and distinguish different phases \cite{Holzhey:1994we,Vidal:2002rm,
Calabrese:2004eu,Kitaev:2005dm,Levin:2006zz}. Moreover, the entanglement entropy enjoys useful geometrical interpretations such as the area law \cite{Bombelli:1986rw,Srednicki:1993im,Eisert:2008ur}, which was motivated to understand the black hole entropy, and also the holographic formula \cite{Ryu:2006bv,Ryu:2006ef,Hubeny:2007xt,Nishioka:2009un}, which allows us to compute the entanglement entropy as the area of a minimal surface in a holographic spacetime.

However, we need to keep in mind that the entanglement entropy measures the amount of quantum entanglement between two subsystems only when the total system is in a pure state \cite{Nishioka:2018khk,Horodecki:2009zz}. The entanglement measures for mixed states have also been intensively explored and many beautiful results have been obtained in quantum information theory \cite{Holzhey:1994we,Bengtsson:2006rfv}. Nevertheless, it is known that, unlike the entanglement entropy, the entanglement measures for mixed states are not unique at all and that most of them turn out to be difficult to calculate as they involve minimization/maximization over huge spaces \cite{Horodecki:2009zz,Bengtsson:2006rfv}.

A slightly more challenging direction is to understand quantum entanglement between multi-partite systems. To explore this, we need to find a quantity which measures quantum correlations between $\tq$ subsystems for $\tq\geq 3$. Recently, a novel and intriguing set of quantities which are called \emph{multi-invariants} have been introduced \cite{Gadde:2024taa}. These are local unitary invariants which can be defined from some number of copies of the density matrix of a quantum state and the operation of partial trace. To each multi-invariant is associated a finite group symmetry and a number of group elements, one for each party. These group elements provide the instructions for how to take the partial traces.

One such example of particular interest is called the multi-entropy and was first considered in \cite{2022PhRvD.106l6001G,Penington:2022dhr}. The $n$-th Rényi $\tq$-partite multi-entropy, denoted by $S^{(\tq)}_n$, is defined from $n^{\tq-1}$ copies of the density matrix of a given $\tq$-partite system by making the contraction of the indices following a systematical rule, which can be regarded as a generalization of replica method of the $n$-th Rényi entanglement entropy.  Therefore, the multi-entropy is a natural generalization of entanglement entropy and can be computed straightforwardly for any quantum many-body systems. The multi-entropy has another remarkable feature that it might have a simple holographic formula in terms of the sum of minimal surfaces generalizing the holographic entanglement entropy \cite{Ryu:2006bv,Ryu:2006ef,Hubeny:2007xt,Nishioka:2009un} as argued in \cite{PhysRevD.108.054508}, though we need to be careful about the replica symmetry breaking as pointed out in \cite{Penington:2022dhr}. 
Several fundamental properties of multi-entropy have been worked out in \cite{Gadde:2023zni,Gadde:2023zzj,Gadde:2024jfi}. The calculations of multi-entropy in two-dimensional conformal field theories (CFTs) have been done by extending the replica method in \cite{Harper:2024ker,Gadde:2024taa}. 
The application of multi-entropy to topological phases was analyzed in \cite{Liu:2024ulq}. The use of multi-entropy in black hole information problem was studied in \cite{Iizuka:2024pzm,Iizuka:2025ioc,Iizuka:2025bcc,Iizuka:2025caq}. An extension of multi-entropy to canonical purification was studied in \cite{Yuan:2024yfg}. 

Another class of measures we will consider are the dihedral measures first defined in  \cite{Gadde:2024taa}. The dihedral measures constitute an infinite family of multi-invariants on 3 parties with $2n$ copies of the density matrix. The group elements defining the measures are taken from the dihedral group of order $2n$. For 2d CFTs they have the exceptional property that for each value of $n$ the replica surface is genus $g=0$. A calculation of the dihedral measures for 2d CFTs was carried out in \cite{Gadde:2024taa}.

The main purpose of this paper is to perform the direct numerical calculations of multi-entropy and the dihedral measures at quantum critical points as a first step to employ multi-invariants as a tool to investigate multi-partite entanglement in quantum many-body systems.  In particular, we will focus on tripartite systems $\tq=3$ in the free scalar field theory on a lattice and the transverse-field Ising model, both in $(1+1)$ dimension. One intriguing quantity is the excess $\kappa^{(\tq)}_n$ defined by the difference between the multi-entropy and averaged sum of the Rényi entanglement entropy as it is UV finite and is expected to characterize the amount of truly tripartite entanglement. We will evaluate this quantity via the Gaussian method for the free scalar field model and via the matrix product state method for the Ising model. 

For the Ising spin model, we will calculate the multi-entropy for $n=2$ and $\tq=3$. Our numerical value of the excess $\kappa^{(3)}_2$ agrees with the general prediction 
\ba
\kappa^{(3)}_2=\frac{c}{4}\log 2,  \label{generalkab}
\ea
from the CFT analysis \cite{Harper:2024ker,Gadde:2024taa}, by setting $c=\frac{1}{2}$. We also numerically calculated the dihedral measure for $n=3$ and successfully reproduced the CFT result. Furthermore, we calculated the multi-entropy when the subsystems are disconnected as a function of the distance between the two subsystems. We again find a reasonable matching with results from the free fermion CFT.  

In the massless free scalar field model, we will find that the excess is vanishing for $n=2$. At first this is surprising because it does not agree with the general CFT result (\ref{generalkab}). However, we will explain why we have $\kappa^{(3)}_2=0$ from the continuum limit calculations based on the conformal map analysis of the $c=1$ CFT. We find that this anomalous behavior is due to the zero mode contributions. Moreover, our numerical calculations give predictions to $n=3$ and $n=4$, including the case with disconnected subsystems, which has never been computed from the CFT method. We will also obtain similar results for the dihedral measures.

This paper is organized as follows. In section II, we review the basic definition and properties of multi-entropy and dihedral measure and explain results from the replica method calculations in two dimensional CFTs. In section III we show the numerical results of multi-entropy and dihedral measure in the free scalar field theory on a lattice in $(1+1)$ dimension.  In section IV, we present results in the one-dimensional transverse-field quantum Ising model. In section V, we summarize the conclusions and discuss future problems. In appendix A, we present useful identities of theta functions. In appendix B, we explain the details of our procedure of the numerical fittings for the transverse field Ising model.

\section{General CFT results}

\subsection{Multi-invariants}
Consider a $\tq$ party pure state \(\lvert \Psi \rangle\) of a general quantum system. We can expand the state in the basis
\begin{equation}
    \lvert \Psi \rangle = \sum_{i_1=1}^{d_1} \ldots \sum_{i_\tq=1}^{d_\tq} \psi_{i_1\ldots i_\tq} ~\lvert e^{i_1}_1\rangle \otimes \ldots \otimes \lvert e^{i_\tq}_\tq \rangle.
\end{equation}
and from this we can form the density matrix $\rho=\lvert \Psi \rangle \langle\Psi \rvert$. We will be interested in a class of local unitary invariant measures or \emph{multi-invariants} \cite{Gadde:2024taa}. These are formed by taking some number of copies of the density matrix and performing contractions of the various parties with respect to a set of permutations.

For a measure on $s$ copies we choose a group of order $|G|=s$ and for each party pick a group element. Each of these group elements is then represented as a permutation forming the regular representation which acts freely and transitively\footnote{That is the only element whose permutation representation has fixed points (cycles of length one) is the identity $e$ and given two permutations $x,y$ there exists a third $g$ such that $y=gx$.}. Each such permutation consist of some number of cycles $m$ with the same length $l$ such that $s=lm$ and the order of the element is $l$. In this way the group $G$ and be viewed as a subgroup of the symmetric group on $s$ elements $S_{s}$.

The multi-invariant corresponding to the choice $(g_1,\ldots, g_\tq)$ of permutation elements is then given as
\begin{align}\label{inv-def}
    &\CE(g_1,\ldots, g_\tq)\notag\\
    &=  (\psi_{i^{(1)}_1\ldots i^{(1)}_\tq} \ldots \psi_{i^{(s)}_1 \ldots  i^{(s)}_\tq}) (\bar \psi^{j^{(1)}_1\ldots j^{(1)}_\tq} \ldots \bar \psi^{j^{(s)}_1 \ldots j^{(s)}_\tq}) \notag\\
    &\ \ \ \ \cdot\delta^{\vec i_1}_{g_1 \cdot \vec j_1} \ldots \delta^{\vec i_\tq}_{g_\tq \cdot \vec j_\tq},\notag\\
  &  {\rm where}, \quad \delta^{\vec i_\tp}_{g_\tp \cdot \vec j_\tp} \equiv \delta^{i_\tp^{(1)}}_{j_\tp^{(g_\tp \cdot 1)}} \ldots \delta^{i_\tp^{(s)}}_{j_\tp^{(g_\tp \cdot s)}}.
\end{align}
where the contractions are performed "with respect to" the permutations that is if $...ij....$ occurs in the permutation of a party then the lower index of copy $i$ for that party is contracted with the upper index of copy $j$ for that party.

In general each choice of permutations does not lead to a unique multi-invariant there as there is a redundancy to act on all of the permutations by left multiplication:
\be \CE(g_1,\ldots, g_\tq) = \CE(g \cdot g_1,\ldots, g \cdot g_\tq),\qquad {\rm for}\quad g\in S_{s}
\ee
This is the freedom to label the $\psi$'s once the labeling of $\bar \psi$'s has been fixed and can be used to set one of the chosen permutations to be the identity $e$. This is the same as taking the partial trace of the density matrix with respect to one of the parties and instead working with the reduced density matrix on $\tq-1$ parties.

While these measures can be calculated for any quantum system, we will mainly be interested in the application to the quantum critical points which correspond to 2d CFTs in this article. Taking $s$ copies of our 2d CFT we can divide them in half at euclidean time $t_e=0$ and partition the entire resulting boundary into $p$ intervals and assign to each a party. In general we allow parties to consist of regions which are the union of multiple intervals. This construction forms $s$ copies of the density matrix. The permutations for a multi-invariants give instructions for how to attach the boundaries of the $s$ copies together. If $...ij....$ occurs in the permutation of a party then the upper half of copy $i$ is attached to the lower half of copy $j$ along the corresponding interval(s) at the boundary. The result is a branched cover.

There is considerable technology available for the computation of measures corresponding to such branched covers. In particular a branched cover can always be uniformized. That is there exist a conformal map from the upper half plane to polygonal region with a number of sides $p$ corresponding to the number of interval of the boundary at $t_e=0$. The angles of the polygonal region is fixed by the number of copies one traverses as one circles the common end point of two interval on the branched cover. This information can easily be determined from the permutations. Consider two adjacent intervals $i,j$ with group elements $g_i$ and $g_j$ then to rotate around the common end point of the two intervals is to first reflect from the upper half plane to a different copy using $g_i$ and then to reflect back across $j$ from the lower half plane to the upper half plane using $g_j^{-1}$. The order of $g_ig_j^{-1}$ tells us how many times we must circle around the end point before returning to the original copy. If $g_ig_j^{-1}$ is of order $l$ then under the uniformizing conformal map this will correspond to an angle of $\frac{\pi}{l}$ in the resulting polygonal region.

The final result of the uniformizing map is a Riemann surface of genus $g$ which is tessellated by the polygonal regions. A consequence of the Riemann-Hurwitz formula is that the genus is completely fixed just from the choice of permutations
\be\label{eq:RH}
2-2g=s\left(2-p+\sum_{i=1}^p\frac{1}{l_i}\right)
\ee
here $|G|=s$, the number of end points of the intervals is $p$, and in the sum $l_i$ is the order of the group element $g_ig_j^{-1}$ for each end point of the intervals.

Supposing one knows the partition function of the genus $g$ Riemann surface $\Sigma$ for the 2d CFT of interest it is then possible to calculate the measure by considering now a conformal map $\Gamma(z)$ from the Riemann surface to the original branched cover. Under such a conformal transformation the partition function is related to the original via the Liouville action
\be
\begin{split}
e^{S_L}\frac{Z_{\Sigma}}{Z^s_{\mathbb{CP}}},\quad S_L=\frac{c}{96\pi}\int_{\Sigma}d^2z\sqrt{g}\left[\partial_\mu\phi\partial_\nu\phi g^{\mu\nu}+2R\phi\right] \\
\phi(z)=2\log\left(|\partial_z\Gamma(z)|\right).
\end{split}
\ee
where $Z_{\mathbb{CP}}$ is the partition function of the Riemann sphere and $R$ is the Ricci curvature of the Riemann surface $\Sigma$.

The Liouville action is divergent and care must be taken when evaluating to properly regulate it. While this procedure works for any genus typically explicit calculations are only performed in the cases $g=0,1$ where the partition functions and maps $\Gamma(z)$ are explicitly known.

Going through the calculation of such measures one can prove that they will take the form of $p$-point functions of conformal primaries. To make this connection more precise one can consider a single copy of the original theory with the operator content duplicated $s$ times and inserts a conformal primary operator called a twist operator. Twist operators $\sigma_{h}(x)$ are labeled by the point of insertion $x$ as well as a monodromy, a permutation element $h$ enacting boundary conditions between the fields. The idea is to choose the boundary conditions identically to that of the action of circling around one of the end points of the branched cover. That is the monodromies of the twist operators are taken to be precisely the same as the group elements $h=g_ig_j^{-1}$. In this way the multi-invariant can also be computed as a $p$-point function of conformal primary twist operators in the replicated theory. The conformal dimension of a twist operator is given by
\be
\Delta_{\sigma_{g}}=\frac{cs}{12}\frac{l^2-1}{l^2}
\ee
where $g$ is a permutation of $m$ cycles of length $l$ such that $s=ml$.

This entire procedure can be seen as the natural generalization of the replica method used for the calculation of holographic (Rényi) entanglement entropies to multi-invariants with more than two parties.

Next we will describe some examples of explicit multi-invariants which will appear in this paper.
\subsubsection{Rényi-entropy with $\tq=p=2$}
The Rényi entropy is given by
\be
S^{(2)}_n(A)=\frac{1}{1-n}\log (\Tr\rho_A^n)
\ee
while the entanglement entropy
\be
S=-\Tr(\rho_A \log \rho_A) = \lim_{n\rightarrow 1}S_n.
\ee

The Rényi entropy can be re-expressed as a multi-invariant with $g$'s taken from the cyclic group $\mathbb{Z}_n$:
\be
\mathbb{Z}_{n}:\langle a|a^n=e\rangle.
\ee
For the generator we choose the representation
\be
   a:\quad (1,2,\cdots,n)
\ee
and make the choice
\be
g_O=e,\quad g_A=a^{n-1}.
\ee
This corresponds to twist operators with monodromy\footnote{In the literature these are typically written in terms of the cycle length as $\sigma_n$ and $\sigma_{\bar{n}}$. This is because the monodromies are always chosen to be of order $n$ taken from group $\mathbb{Z}_n$. As we generalize to more parties and other group symmetry the permutations will contain multiple cycles and it is more informative to label the twist operators by the group element rather than the order.}
\be
\begin{split}
   \sigma_{a}:&\quad (1,2,\cdots, n)\\ 
   \sigma_{a^{n-1}}:&\quad  (1,n,n-1,\cdots,2).
\end{split}
\ee

For example consider the case $n=2$ then action of $g_O$ is equivalent to taking the partial trace over $O$ for each copy of the density matrix
\be
\rho_A=\psi_{ia}\bar\psi^{ja}
\ee
and
\be
\CE(e,a^{n-1})=\Tr\rho_A^2=\psi_{ia}\psi_{jb}\bar\psi^{ja}\bar\psi^{ib}
\ee
so that
\be
S^{(2)}_2(A)=-\log(\CE(e,a^{n-1}))
\ee

For a single interval $A=[x_1,x_2]$ of a 2d CFT the n-th Rényi entropy can be found from
\ba
(1-n)S^{(2)}_n(A)&=&\log\langle\sigma_a(x_1)\sigma_{a^{n-1}}(x_2)\rangle\no
&=&-\frac{c}{6}\frac{n^2-1}{n}\log|x_2-x_1|
\ea
This leads to the expression of Rényi entropy:
\ba
S^{(2)}_n(A)=\frac{c}{6}\left(1+\frac{1}{n}\right)\log\frac{|x_2-x_1|}{\ep}, \label{REEa}
\ea
where we recovered the dependence on the UV cut off $\ep$\footnote{In regulating the Liouville action one must choose a scale for which the two point function of twist operators $\langle\sigma_g(x_i)\sigma_{g^{-1}}(x_j)\rangle=1$. Typically one chooses this scale to be $|x_i-x_j|=1$. If one instead sets this scale to be $\ep$ this restores the usual UV cut off. In practice one can use dimensional analysis to restore the UV cut off if so desired.}.

\subsubsection{Three party multi-entropy $\tq=p=3$}
An important family of measures the \emph{Rényi multi-entropy} was first defined in \cite{2022PhRvD.106l6001G,Penington:2022dhr}. These measures give rise to twist operator monodromies where the cycle length of all operators is the same. For the case of three parties $\tq=3$ we have that the n-th Rényi multi-entropy is defined by
\be
\mathbb{Z}^2_{n}:\langle a,b|a^n=b^n=e;ab=ba\rangle
\ee
\be
\begin{split}
   a:&\quad (1,2,\cdots, n)(n+1,n+2,\cdots, 2n)\cdots\\
   &\cdots ((n-1)n+1,(n-1)n+2,\cdots,n^2),\\
   b:&\quad  (1,n+1,\cdots,(n-1)n+1))(2,n+2,\cdots,\\
   &\cdots (n-1)n+2)\cdots(n,2n,\cdots,n^2).\\
\end{split}
\ee
\be\label{eq:me3perm}
g_O=e,
\;g_A=a^{n-1},
\;g_B=b
\ee
For example if $n=2$
\be
\CE(e,a^{n-1},b)=\psi_{iwa}\psi_{jxb}\psi_{kyc}\psi_{lzd}\bar\psi^{jya}\bar\psi^{izb}\bar\psi^{lwc}\bar\psi^{kxd}
\ee
This gives the twist operator monodromies
\be
\begin{split}
   \sigma_{a}:&\quad (1,2,\cdots, n)(n+1,n+2,\cdots, 2n)\cdots\\
  & \cdots((n-1)n+1,(n-1)n+2,\cdots,n^2))\\ 
   \sigma_{a^{n-1}b^{n-1}}:&\quad  (1,n^2,\cdots n+2),(2,(n-1)n+1,\cdots n+3)\cdots\\
  &\cdots (n,n^2-1,\cdots n+1)\\
   \sigma_{b}:&\quad  (1,n+1,\cdots,(n-1)n+1))\\
   &\ \ \ \ (2,n+2,\cdots,(n-1)n+2)\cdots\\
  &\ \ \ \  (n,2n,\cdots,n^2).
\end{split}
\ee
From this we chose a normalization and define the multi-entropy to be\footnote{Though we will not make further use of it in this paper the multi-entropy can be defined for any number of parties $\tq$. The permutations are taken from the group $\mathbb{Z}_n^{\tq-1}$ such that the corresponding monodromies of the twist operators each consist of $n^{\tq-2}$ cycles of length $n$. The normalization is then taken to be $
S_n^{(\tq)}=\frac{1}{1-n}\frac{1}{n^{\tq-2}}\log\mathcal{E}$.}
\be\label{eq:MEq3}
S^{(3)}_{n}=\frac{1}{1-n}\frac{1}{n}\log(\CE(e,a^{n-1},b))
\ee

We choose $A=[x_1,x_2]$ and $B=[x_2,x_3]$ in a 2d CFT. In the replica method calculation, the conformal symmetry fixes the multi-entropy to be the form
\ba\label{eq:Snq3}
S^{(3)}_n=\frac{c}{12}\left(1+\frac{1}{n}\right)\log \frac{x_{21}x_{31}x_{32}}{\ep^3} +\kappa^{(3)}_n,
\ea
where $x_{ij}=x_i-x_j$. Here the choice of the UV cut off $\ep$ is fixed by (\ref{REEa}) and thus the constant part  $\kappa_n$ is not ambiguous. In order to calculate $\kappa_n$, we need to explicitly evaluate the replicated partition function including the correct normalization. For $n=2$, this was evaluated in \cite{Harper:2024ker,Gadde:2024taa} as in (\ref{generalkab}) i.e.  $\kappa^{(3)}_2=\frac{c}{4}\log 2$. If we assume the conjectured holographic formula of \cite{2022PhRvD.106l6001G}, this leads to the prediction $\kappa^{(3)}_1=\frac{c}{2}\log \frac{2}{\s{3}}$. Notice also that this constant coincides with the difference between the multi-entropy and the averaged entanglement entropies:
\be
\Delta S^{(3)}_n=S_n^{(3)}-\frac{1}{2}\left(S^{(2)}_n (A)+S^{(2)}_n (B)+S^{(2)}_n (O)\right), \label{difme}
\ee
where the UV cut off dependence cancels. 

\subsubsection{Three party multi-entropy on four intervals $\tq=3$ and $p=4$}
We choose $A=[x_1,x_2]$ and $B=[x_3,x_4]$ and $O=(A\cup B)^c$ and define the cross-ratio
\be
\eta=\frac{x_{12}x_{34}}{x_{13}x_{24}},\label{crossr}
\ee
where $x_{ij}=x_i-x_j$.
Taking the same permutations \eqref{eq:me3perm} gives the twist operator mondromies

\be
\begin{split}
   \sigma_{a}:&\quad (1,2,\cdots, n)(n+1,n+2,\cdots, 2n)\cdots\\
  & \cdots((n-1)n+1,(n-1)n+2,\cdots,n^2))\\ 
    \sigma_{a^{n-1}}:&\quad (1,n,\cdots, 2)(n+1,2n,\cdots, n+2)\cdots\\
  & \cdots((n-1)n+1,n^2,\cdots,(n-1)n+2))\\ 
   \sigma_{b^{n-1}}:&\quad  (1,(n-1)n+1),\cdots,n+1)\\
   &\ \ \ \ (2,(n-1)n+2,\cdots,n+2)\cdots\\
  &\ \ \ \  (n,n^2,\cdots,2n)\\
    \sigma_{b}:&\quad  (1,n+1,\cdots,(n-1)n+1))\\
   &\ \ \ \ (2,n+2,\cdots,(n-1)n+2)\cdots\\
  &\ \ \ \  (n,2n,\cdots,n^2).
\end{split}
\ee

In the case $n=2$ this can be explicitly evaluated for a 2d CFT
\ba
\begin{split}
&\log(\langle\sigma(0)_{g_0}\sigma(\eta)_{g_\eta}\sigma(1)_{g_1}\sigma(\infty)_{g_\infty}\rangle)=\\
&-\frac{c}{6}\log(\eta(1-\eta))-\frac{4c}{3}\log2+\log(Z_{\text{torus}}(\tau))
\end{split}
\ea
which is theory dependent due to the presence of the torus partition function. The modular parameter of the torus is pure imaginary $\tau=i|\tau|$ and is related to the cross-ratio by the relation
\be
\eta=\left(\frac{\theta_2(\frac{\tau}{2})}{\theta_3(\frac{\tau}{2})}\right)^4.
\ee

This leads to the following expression of the multi-entropy \cite{Harper:2024ker}:
\ba
 S^{(3)}_2 &=&\frac{2c}{3}\log 2+\frac{c}{12}\log \frac{x_{12}x_{13}x_{14}x_{23}x_{24}x_{34}}{\ep^6}\no
&&\ \  -\frac{1}{2}\log Z_{\text{torus}}(\tau).\label{Sthre}
\ea

\subsubsection{Three party dihedral measures $\tq=p=3$}
An infinite family of three party measures with genus 0 replica surface was first defined in \cite{Gadde:2024taa}. These measures are made of $2n$ copies with dihedral replica symmetry
\be
\mathbb{D}_{2n}:\langle a,b|a^n=b^2=e;ba=a^{n-1}b\rangle
\ee
\be
\begin{split}
   &a:\quad (1,2,\cdots, n)(n+1,n+2,\cdots, 2n)\\
   &b:\quad  (1,2n)(2,2n-1)(3,2n-2)\cdots(n,n+1)\\.
\end{split}
\ee
\be
g_O=e,
\;g_A=a^{n-1},
\;g_B={b}
\ee
which gives the twist operator monodromies
\be
\begin{split}
   \sigma_{a}:&\quad (1,2,\cdots, n)(n+1,n+2,\cdots, 2n)\\ 
   \sigma_{a^{n-1}b}:&\quad  (1,n+1)(2,2n)(3,2n-1)\cdots(n,n+2)\\
   \sigma_{b}:&\quad  (1,2n)(2,2n-1)(3,2n-2)\cdots(n,n+1).
\end{split}
\ee
The three point function for 2d CFT was found to be
\be\label{eq:full3ptdi}
\begin{split}
&\log(\langle\sigma(x_1)_a\sigma(x_2)_{a^{n-1}b}\sigma(x_3)_b\rangle)\\
&=-\frac{c}{6}\frac{n^2-1}{n} \log(|x_2-x_1|)-\frac{c}{6}\frac{n^2-1}{n} \log(|x_3-x_1|)\\
&\ \ \ -\frac{c}{12}\frac{n^2+2}{n} \log(|x_3-x_2|)-\frac{c}{3}\frac{n^2-1}{n}\log 2.
\end{split}
\ee
Choosing normalization we define the dihedral three party measure to be
\be
\mathcal{D}_{2n}=\frac{1}{1-n}\frac{1}{2n}\log(\langle\sigma(x_1)_a\sigma(x_2)_{a^{n-1}b}\sigma(x_3)_b\rangle)
\ee
which suggests that we consider the excess
\ba
\Delta\mathcal{D}_{2n} &=\mathcal{D}_{2n}-\frac{1}{2n}\Bigl(S_n^{(2)}(A)+S_n^{(2)}(B)\\
 &+\frac{1}{2}\frac{n^2+2}{n^2-1}S_n^{(2)}(O)\Bigr).
\ea

For 2d CFTs this reduces to
\be
\Delta\mathcal{D}_{2n}=\frac{c}{6}\frac{n+1}{n^2}\log(2).
\label{diheu}
\ee

\subsection{Basic behaviors of the multi-entropy and dihedral measures in qubit systems}
We investigate the multi-entropy and dihedral measures in two representative classes of three-qubit pure states with tripartite entanglement: the generalized GHZ states and the generalized W states.
In addition, we examine the Werner state, which arises as a reduced density matrix of a pure state in a larger Hilbert space.

\subsubsection{GHZ states}
\begin{figure*}[tbp]
    \centering
    \includegraphics[width=0.45\textwidth]{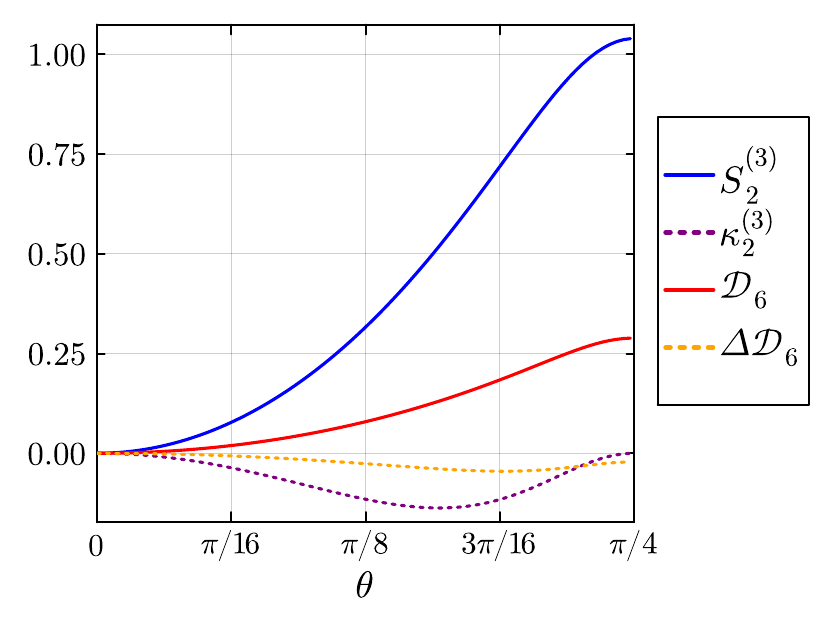}
    \includegraphics[width=0.45\textwidth]{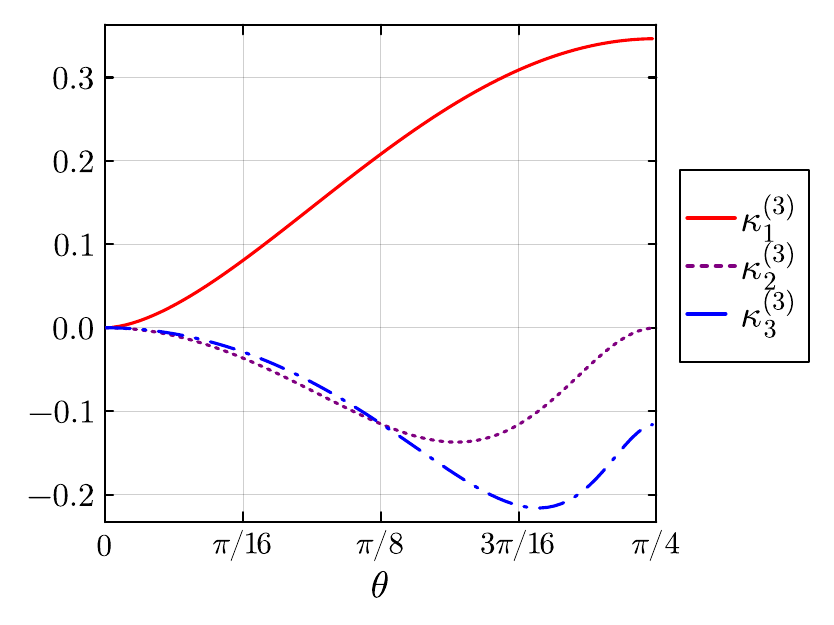}
    \caption{
        Multi-entropy of the generalized GHZ state as functions of $\theta$
        \textit{Left:}~The three-party multi-entropy $S_2^{(3)}$, the dihedral measure $\mathcal{D}_6$, and their excess.
        \textit{Right:}~The multi-entropy excess $\kappa_n^{(3)}$ for $n=1, 2, 3$.
    }
    \label{fig:GHZ}
\end{figure*}

The generalized GHZ state is defined in the computational basis as
\begin{equation}
    \ket{\text{GHZ}(\theta)} \coloneqq \cos{\theta} \ket{000} + \sin{\theta} \ket{111}.
\end{equation}
For this family, the three-party $n$-th multi-entropy and the single-qubit Rényi entropy are analytically given by
\begin{align}
    S_n^{(3)} &= \frac{1}{(1-n)n} \log[(\cos{\theta})^{2n^2} + (\sin{\theta})^{2n^2}],\\
    S_n^{(2)} &= \frac{1}{1-n} \log[(\cos{\theta})^{2n} + (\sin{\theta})^{2n}].
\end{align}
In the von Neumann limit $n\to 1$, both quantities converge to the same expression as
\begin{equation}
    \frac{1}{2} S_1^{(3)} = S_1^{(2)} = - \cos^2{\theta}\log[\cos^2{\theta}] - \sin^2{\theta} \log[\sin^2{\theta}].
\end{equation}
The multi-entropy exceses, for example for $n=2$ and $n=1$, are respectively given by
\begin{align}
    \kappa_2^{(3)} &= \frac{1}{2} \log\left[ \frac{(\cos^4{\theta}+ \sin^4{\theta})^3}{\cos^8{\theta} + \sin^8{\theta}} \right]\\
    \kappa_1^{(3)} &= - \frac{1}{2} \cos^2{\theta} \log[\cos^2{\theta}] - \frac{1}{2} \sin^2{\theta} \log[\sin^2{\theta}].
\end{align}
Although the multi-entropy excess is non-positive for finite Rényi index, its von Neumann limit $\kappa_1^{(3)}$ is non-negative, as shown in the right panel of Fig.~\ref{fig:GHZ}.
The dihedral measure is given by
\begin{equation}
    \mathcal{D}_{6} = -\frac{1}{12} \log\left[ \cos^{12}{\theta} + \sin^{12}{\theta} \right].
\end{equation}

\subsubsection{W states}
\begin{figure*}[tbp]
    \centering
    \includegraphics[width=0.45\textwidth]{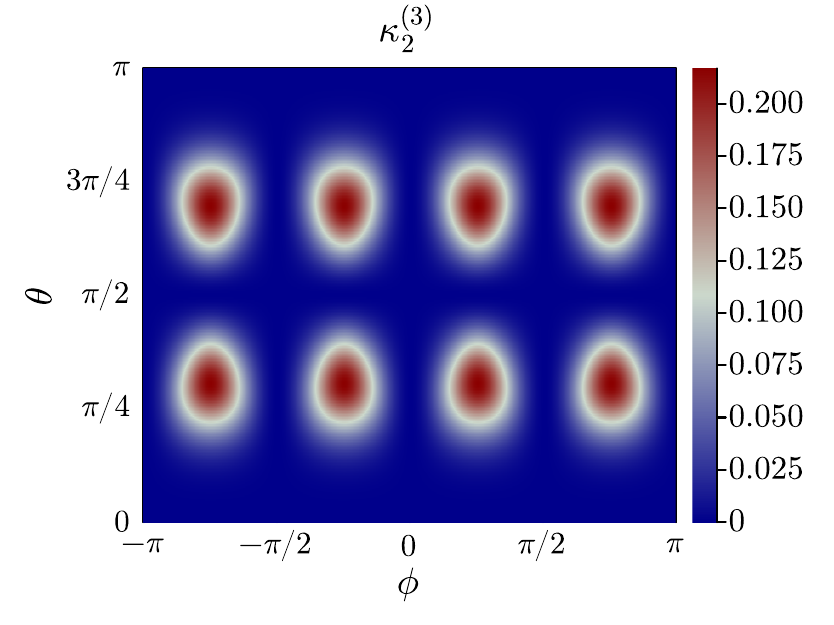}
    \includegraphics[width=0.45\textwidth]{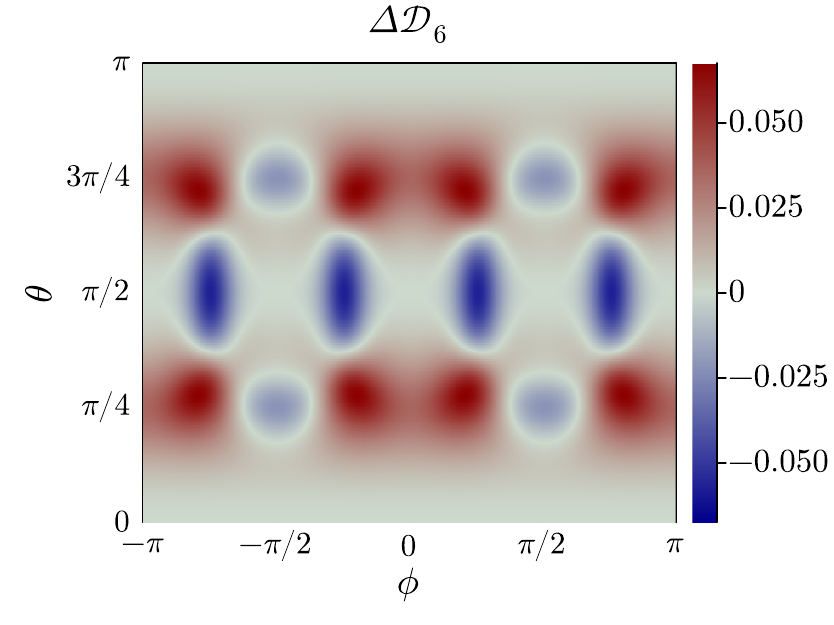}
    \caption{
        Excess for the generalized W state, plotted as functions of $\theta$ and $\varphi$.
        \textit{Left:}~The multi-entropy excess $\kappa_2^{(3)}$.
        \textit{Right:}~The dihedral excess $\Delta\mathcal{D}_6$.
    }
    \label{fig:W}
\end{figure*}

The generalized W state is defined in the computational basis as
\begin{equation}
    \ket{\text{W}(\theta, \varphi)} \coloneqq \cos{\theta} \ket{001} + \sin{\theta} \cos{\varphi} \ket{010} + \sin{\theta} \sin{\varphi} \ket{100}.
\end{equation}
For this family, the multi-entropy and the single-qubit Rényi entropy are given by
\begin{align}
    S_2^{(3)} &= - \log\left[ \cos^4{\theta} + \frac{3 + \cos{4\varphi}}{4} \sin^4{\theta} \right],\\
    S_n^{(2)}(A) &= \frac{1}{1-n} \log[ (\cos{\theta})^{2n} + (\sin{\theta})^{2n} ],\\
    S_n^{(2)}(B) &= \frac{1}{1-n} \log[ (\sin^2{\theta} \cos^2{\varphi})^n + (\cos^2{\theta} + \sin^2{\theta} \sin^2{\varphi})^n ],\\
    S_n^{(2)}(C) &= \frac{1}{1-n} \log[ (\sin^2{\theta} \sin^2{\varphi})^n + (\cos^2{\theta} + \sin^2{\theta} \cos^2{\varphi})^n ].
\end{align}
For the dihedral measure, we have
\begin{widetext}
\begin{equation}
    \mathcal{D}_6 = -\frac{1}{12} \log\left[ \left(\cos^4{\theta} + \frac{3+\cos{4\varphi}}{4}\sin^4{\theta} - \cos^2{\theta}\sin^2{\theta}\cos^2{\varphi} \right)^2 \left( \cos^4{\theta} + \frac{3+\cos{4\varphi}}{4}\sin^4{\theta} + 2 \cos^2{\theta}\sin^2{\theta}\cos^2{\varphi} \right) \right].
\end{equation}
\end{widetext}
The behavior of these quantities is illustrated in Fig.~\ref{fig:W}.

\subsubsection{Werner states}
\begin{figure}[tbp]
    \centering
    \includegraphics[width=0.45\textwidth]{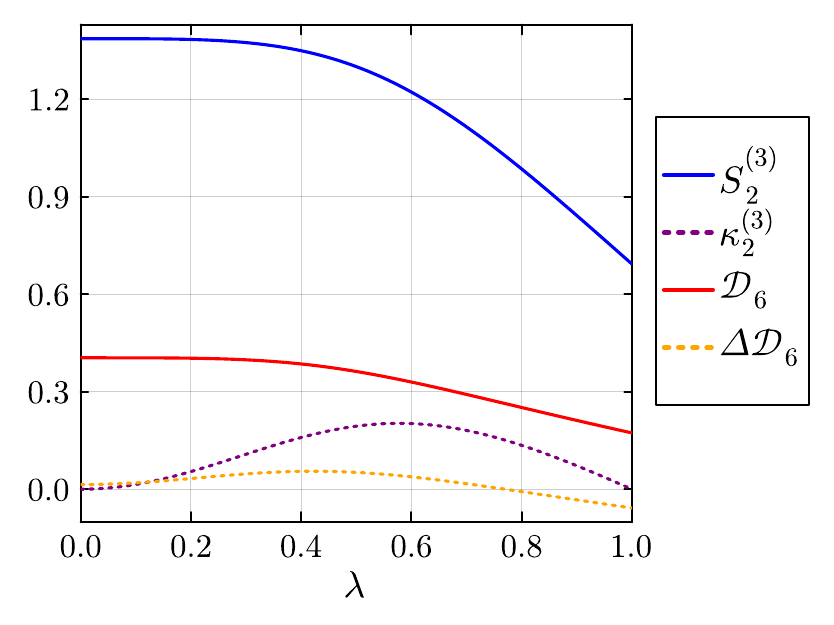}
    \caption{
        Multi-entropy and dihedral measure of the Werner state.
    }
    \label{fig:Werner}
\end{figure}

The two-qubit Werner states are a family of mixed states that is a convex combination of a Bell state and the maximally mixed state, defined as
\begin{align}
    \rho_{\text{Werner}}(\lambda) &\coloneqq \lambda \dyad{\Psi^-}{\Psi^-}_{AB} + \frac{1-\lambda}{4} \openone_{AB},\\
    \ket{\Psi^-} &\coloneqq \frac{1}{\sqrt{2}} (\ket{01} - \ket{10}),
\end{align}
where $\lambda$ is a parameter determining the amount of entanglement in the state.
We consider a scenario where the Werner state arises as a two-qubit reduced density matrix of a pure state on a larger Hilbert space. 
The multi-entropy and Rényi entropy for the original pure state are given by
\begin{align}
    S_2^{(3)} &= - \frac{1}{2} \log\left[ \frac{3\lambda^4 + 1}{16} \right],\\
    S_n^{(2)}(O) &= \frac{1}{1-n} \log\left[ 3 \left(\frac{1-\lambda}{4}\right)^n + \left( \frac{3\lambda+1}{4} \right)^n \right],\\
    S_n^{(2)}(A) &= S_n^{(2)}(B) = \log{2}.
\end{align}
For the dihedral measure, we have
\begin{equation}
    \mathcal{D}_6 = -\frac{1}{12} \log\left[ \frac{6\lambda^6 + 9\lambda^4 + 1}{128} \right].
\end{equation}
These quantities are plotted in Fig.~\ref{fig:Werner}.

\subsection{More analysis of three party multi-entropy on four intervals}
\label{sec:multi-dis}
It is useful to examine the behavior of the multi-entropy in 2d CFTs more explicitly. In particular we focus on the case of the three party multi-entropy on four intervals given by the general formula (\ref{Sthre}) as we will perform numerical lattice calculations later for this in the free scalar and Ising model. 

It is also helpful to consider the second Rényi entropies
defined by (\ref{REEa}) at $n=2$. First of all, for the subsystems $A$ and $B$, we immediately find
 \be
S^{(2)}_2(A)=\frac{c}{4}\log \frac{x_{21}}{\ep},\ \ \ \ 
S^{(2)}_2(B)=\frac{c}{4}\log \frac{x_{43}}{\ep}.
 \ee
For the subsystem $AB$, which is the union of two disconnected subsystems, is computed via the standard replica method as follows: 
\ba
S^{(2)}_2(AB)&=&\frac{2c}{3}\log 2 + \frac{c}{12}\log \frac{x_{12}x_{13}x_{14}x_{23}x_{24}x_{34}}{\ep^6} \no
&& - \log \left[Z_{\text{torus}}\left(\frac{\tau}{2}\right)\right],
\ea
where $\tau$ is related to $x_i$ via  (\ref{crossr}). Thus by combining this with (\ref{Sthre}), the excess 
(\ref{difme}) is expressed as 
\ba
\Delta S^{(3)}_2&=&\frac{c}{3}\log 2 +\frac{c}{24}\log \frac{x_{13}x_{14}x_{23}x_{24}}{x^2_{12}x^2_{34}}\no
&&-\frac{1}{2}\log Z_{torus}(\tau)+\frac{1}{2}\log Z_{torus}\left(\frac{\tau}{2}\right).\label{difmee}
\ea

In the  limit $x_2\to x_3$, we find
\ba
&& S^{(3)}_2\simeq \frac{c}{2}\log 2 +\frac{c}{8}\log \frac{x_{32}}{\ep}+\frac{c}{8}
\log\frac{x_{21}x_{42}x_{41}}{\ep^3},\label{metaa}\\
&& S^{(2)}_2(AB)\simeq \frac{c}{4}\log\frac{x_{41}}{\ep}+\frac{c}{4}\log\frac{x_{32}}{\ep}.
\ea
We can evaluate $\kappa_2$ by directly computing 
(\ref{difmee}). Naively, this leads to $\Delta S^{(3)}_2=\frac{c}{2}\log 2$. However, as argued in \cite{Harper:2024ker} from a careful analysis, for the multi-entropy $S^{(3)}_2$, we can regard the interval $[x_2,x_3]$ vanishing when $x_3-x_2=\frac{\ep}{4}$, while for the Rényi entropy $S^{(2)}_2$ the interval vanishes at $x_3-x_2=\ep$. This leads to
\ba
\kappa_2^{(3)}= \frac{c}{4}\log 2. \label{devb}
\ea

On the other hand, in the limit $x_1\to x_2$ we obtain
\ba
&& S^{(3)}_2\simeq \frac{c}{4}\log \frac{x_{21}}{\ep}+\frac{c}{4}
\log\frac{x_{43}}{\ep},\label{metbb}\\
&& S^{(2)}_2(AB)\simeq \frac{c}{4}\log\frac{x_{21}}{\ep}+\frac{c}{4}\log\frac{x_{43}}{\ep}.
\ea
Thus we find $\Delta S^{(3)}_2\simeq 0$ in this limit.

\subsubsection{$c=\frac{1}{2}$ Ising CFT}

For the Ising CFT i.e. the Majorana fermion CFT at $c=\frac{1}{2}$, the partition function reads 
\ba
Z_{\text{torus}}(\tau)=\frac{1}{2}\left[\frac{\theta_2(\tau)}{\eta(\tau)}+\frac{\theta_3(\tau)}{\eta(\tau)}+\frac{\theta_4(\tau)}{\eta(\tau)}\right].
\ea
Using the identities (\ref{thetai}), we obtain the analytical expression of the difference (\ref{difme}): 
\ba
&& \Delta S^{(3)}_2=\frac{1}{4}\log 2 +\frac{1}{48}\log \left[\frac{1-\eta}{\eta^2}\right] \no
&&+\frac{1}{2}\log\! \left[\!\frac{\eta^{\frac{1}{12}}(1\!-\!\eta)^{-\frac{1}{24}}
\!+\!\eta^{\frac{1}{3}}(1\!-\!\eta)^{-\frac{1}{24}}
\!+\!\eta^{\frac{1}{12}}(1\!-\!\eta)^{\frac{5}{24}}
}{\s{1\!-\!(1\!-\!\eta)^{\frac{1}{2}}}
\!+\!\s{1+(1\!-\!\eta)^{\frac{1}{2}}}
\!+\!\s{2(1\!-\!\eta)^{\frac{1}{4}}}}\!\right].\no
\label{IsingDS}
\ea
In the limit $\eta\to 0$ we find $\Delta S^{(3)}_2=0$, while in the limit
$\eta\to 1$, we get  $\Delta S^{(3)}_2=\frac{1}{4}\log 2$, which should be interpreted as $\kappa_2=\frac{1}{8}\log 2$, as in (\ref{devb}). 

\begin{figure}[htbp]
        \centering
            \includegraphics[width=6cm]{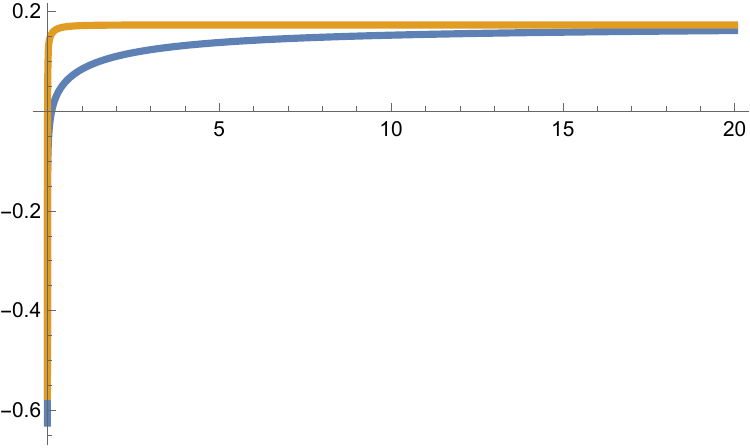}
            \includegraphics[width=6cm]{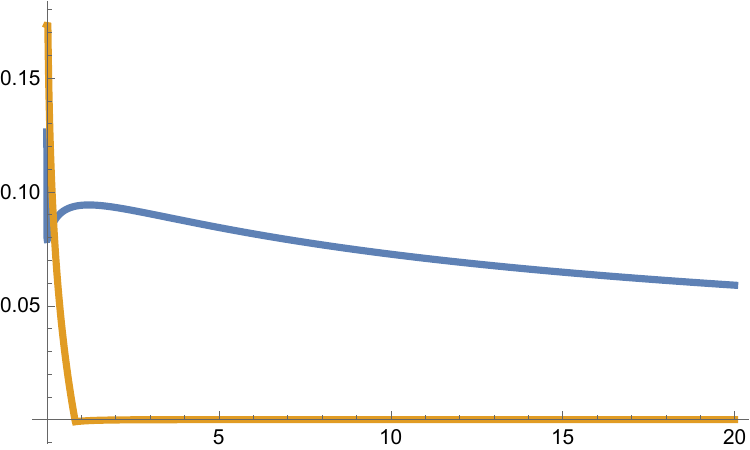}
        \caption{Plots of the multi-entropy $S^{(3)}_2$ (top) and its excess $\Delta S^{(3)}_2$ (bottom) as a function $d$, for the choice of the subsystem $A=[0,2]$ and $B=[2+d,4+d]$.
        The blue and orange graph describe the results for the Ising CFT (\ref{IsingDS}) and the holographic CFT (\ref{holS}), respectively. We set $\ep=1$ and for the latter we choose $c=\frac{1}{2}$.} 
        \label{fig:MENb}
\end{figure}

When we choose $A=[0,2]$ and $B=[2+d,4+d]$, the cross ratio reads $\eta=\frac{4}{(d+2)^2}$. The behaviors of $S^{(3)}_2$ and $\Delta S^{(3)}_2$ are plotted in Fig.\ref{fig:MENb}. For large $d$ we find
 \ba
&& \Delta S^{(3)}_2\simeq \frac{1}{2\s{2}\s{d}}\ \ \ (d\to\infty).\label{eq:IsingCFT_excess_distant}
 \ea
 The behaviors of $S^{(3)}_2$ and 
$ \Delta S^{(3)}_2$ are plotted as the blue graphs in Fig.\ref{fig:MENb}.

\subsubsection{$c=1$ Free scalar}
In the $c=1$ CFT of the massless free scalar, the torus partition function is given by 
\ba
Z_{\text{torus}}(\tau)=\frac{V}{\s{\tau_2}\eta(\tau)^2}
=\frac{V\s{\tau_2}}{\eta\left(-\frac{1}{\tau}\right)^2},\label{torusp}
\ea
where $V$ is the infinite volume of the zero mode.

By employing the identities in (\ref{thetai}),
the excess $\Delta S^{(3)}_2$ turns out to be just a constant:
\ba
\Delta S^{(3)}_2=\frac{1}{4}\log 2.
\ea

If we admit the prescription which fills the deviation between $\Delta S^{(3)}_2$ in the limit $x_2\to x_3$ and $\Delta S^{(3)}_2$ in the coincident $x_2=x_3$ case (\ref{devb}), then we may argue 
\ba\label{eq:deltaSn2Scalar}
\kappa^{(3)}_2=0. \label{fscab}
\ea
Notice that this result is an exception to the general result (\ref{devb}). This arises due to the zero mode of the massless scalar field which leads to the factor $\s{\tau_2}$ in the torus partition function (\ref{torusp}). In the analysis of (\ref{devb}), we assumed the behavior $Z_{torus}(\tau)\simeq e^{\frac{\pi c}{6\tau_2}}$, in the limit $\tau_2\to 0$ and this is correct for any unitary CFTs with the discrete spectrum.  In the free scalar field theory, on the other hand, the integration over the continuous momentum mode gives the extra factor $\s{\tau_2}$ as in (\ref{torusp}). Indeed, later this result (\ref{fscab}) will be confirmed by our numerical results on the lattice soon.

\subsubsection{Holographic CFT}

For a holographic 2d CFT the torus partition function is given by
\be
\log(Z_{\text{torus}})\!=\!
\begin{cases}
        \frac{\pi c}{3}\frac{K(1-\eta)}{K(\eta)}\!=\!-\frac{\pi ic\tau}{6} & \eta\leq \eta_*\\
        \frac{\pi c}{12}\frac{K(\eta)}{K(1-\eta)}\!=\!\frac{\pi ic}{6\tau}  & \eta\geq \eta_*
    \end{cases}, \quad \eta_*\!=\!\frac{\theta_2(\frac{i}{2})^4}{\theta_3(\frac{i}{2})^4},
\ee
the phase transition occurs when the torus is square corresponding to the modular parameter $\tau=i$.
This leads to the following formula for $S^{(3)}_2$:
\ba
&& S^{(3)}_{2}= \no 
&& \left\{
\begin{array}{ll}
\frac{2c}{3}\log 2+\frac{c}{12}\log[x_{12}x_{13}x_{14}x_{23}x_{24}x_{34}] 
-\frac{\pi i c}{12\tau} \  (\eta>\eta_*)\\
 \frac{2c}{3}\log 2+\frac{c}{12}\log[x_{12}x_{13}x_{14}x_{23}x_{24}x_{34}] 
+\frac{\pi i c\tau}{12} \   (\eta<\eta_*).\label{holS}
\end{array}
\right.\no
\label{sthphase}
\ea
The phase transition point is at $\eta_*\simeq 0.97$.
This corresponding to a very small value of $d$ for the choice $A=[0,2]$ and $B=[2+d,4+d]$, given by $d\simeq 0.03$. The behavior is plotted in Fig.\ref{fig:MENb} as the orange graphs. When compared with the Ising CFT, the excess decays very quickly in the holographic case due to the phase transition of the Rényi entropy $S^{(2)}_2(AB)$.


\section{Numerical multi-entropy for 2d Free Scalar Field CFT}
In this section we study multi-entropy in Gaussian states of the one-dimensional harmonic lattice model. We focus on Gaussian states that are special due to their characterization in terms of their covariance matrix that makes the simulation of these states effective since the the covariance matrix size grows linearly with the size of the system.

The $N$-mode harmonic lattice model with periodic boundary condition is defined as 
\be
H=\frac{1}{2}\left(\frac{1}{\epsilon}p^\top\cdot p+q^\top\cdot \mathbf{V}\cdot q\right)
\ee
where $p^\top=(p_1,p_2,\cdots,p_N)$ and $q^\top=(q_1,q_2,\cdots,q_N)$ and $(q_i,p_i)$ denote the canonical variables of the $i$th oscillator and $\mathbf{V}$ is a circulant matrix defined as $\mathbf{V}=\frac{1}{\epsilon}\mathrm{circ}\left(2+\epsilon^2m^2,-1,-1,\cdots,-1\right)$. This model at large-$N$ is identified with 2$d$ free scalar theory regularized on a lattice with $\epsilon$ denoting the lattice spacing and $m$ denoting the mass parameter. Hereafter we set $\epsilon=1$, denoting unit UV cutoff in the free scalar theory language. In order to compare our results with $c=1$ CFT calculations we take the $m\to0$ limit. 

We will use the path integral method to calculate the multi-entropy. Our main focus will be on the ground state which is a Gaussian state. Since the contractions defining multi-entropy are defined on the spatial manifold, we consider the representation of the Gaussian state in the \textit{position} basis. Let us first remind the reader that in the position basis the ground state, and its corresponding density matrix, up to normalization constants are represented as
\begin{align}
\begin{split}
    \Psi &\propto \exp{\left(-\frac{1}{2}q^\top\cdot \mathbf{W}\cdot q\right)}
    \\
    \rho &\propto \exp{\left[-\frac{1}{2}(q^\top\cdot \mathbf{W}\cdot q+{q'}^\top\cdot \mathbf{W}\cdot q')\right]}
\end{split}    
\end{align}
where $\mathbf{W}=\mathbf{V}^{1/2}$, $q$ and $q'$ denote the indices corresponding to the ket and the bra states. In order to represent the reduced density matrix, we consider the following notation: $\alpha,\beta$ indices correspond to the part that will be traced out, and $a,b$ indices correspond to those oscillators remaining in the reduced density matrix. So we consider the following block form
\be
\mathbf{W}=\begin{pmatrix}
    \mathbf{A}_{ab} & \mathbf{B}_{a\beta} \\ \mathbf{B}^{\top}_{\alpha b} & \mathbf{C}_{\alpha\beta}
\end{pmatrix}
\;\;\;,\;\;\;
\mathbf{W}^{-1}=\begin{pmatrix}
    \mathbf{D}_{ab} & \mathbf{E}_{a\beta} \\ \mathbf{E}^{\top}_{\alpha b} & \mathbf{F}_{\alpha\beta}
\end{pmatrix}
\ee
where from $\mathbf{W}\cdot \mathbf{W}^{-1}=\mathbb{1}$ we find
\begin{align}\label{eq:identity}
\begin{split}
\mathbf{A}\cdot \mathbf{D}+\mathbf{B}\cdot \mathbf{E}^{\top}&=\mathbf{C}\cdot \mathbf{F}+\mathbf{B}^{\top}\cdot \mathbf{E}=\mathbb{1}
\\
\mathbf{A}\cdot \mathbf{E}+\mathbf{B}\cdot \mathbf{F}&=\mathbf{C}\cdot \mathbf{E}^{\top}+\mathbf{B}^{\top}\cdot \mathbf{D}=0
\end{split}    
\end{align}
Performing a number of Gaussian integrals leads to the following form for the reduced density matrix
\begin{align}\label{eq:rhored}
\begin{split}
\rho_{\mathrm{red.}}(q,q')
&=
\left(\mathrm{det}\left(\frac{\mathbf{A}+4\mathbf{X}}{\pi}\right)\right)^{\frac{1}{2}}\times
\\&\;\;\;\;\;\;\;\;\;\;\;
\exp{\left[-\frac{1}{2}
\begin{pmatrix}    q^{\top} & {q'}^{\top}\end{pmatrix}
\cdot\mathbf{\mathcal{M}}\cdot
\begin{pmatrix} q \\ {q'}\end{pmatrix}\right]}\,,
\\
\mathbf{\mathcal{M}}&=\begin{pmatrix}    \mathbf{A} + 2\mathbf{X} & 2\mathbf{X} \\ 2\mathbf{X} & \mathbf{A}+2\mathbf{X}\end{pmatrix}\,,
\end{split}    
\end{align}
where we have included the normalization factor for the reduced density matrix and $\mathbf{X}$ is defined as $\mathbf{X}=-\frac{1}{4}\mathbf{B}\cdot \mathbf{C}^{-1}\cdot \mathbf{B}^{\top}$. Note that $q$ and $q'$ vectors carry $a,b$ indices. Using Eq. \eqref{eq:identity} we can show that $\mathbf{B}\cdot \mathbf{C}^{-1}\cdot \mathbf{B}^{\top}=\mathbf{A}-\mathbf{D}^{-1}$ which has the advantage of expressing the reduced density matrix in terms of the two-point correlators of its internal oscillators. Note that in the ground state, the reduced density matrix is expressed in terms of the diagonal blocks of the covariance matrix, namely  $\mathbf{A}_{ab}=2\langle p_a p_b\rangle$ and $\mathbf{D}_{ab}=2\langle q_a q_b\rangle$, where the expectation values are evaluated in the ground state.

\subsection{Case $\tq=2$}
As a warm up we first reproduce the well known Rényi entropies for $\tq=2$ with our approach.  $\mathrm{Tr}\rho^n_{\mathrm{red.}}$ is given by the following path integral 
\begin{align}
\begin{split}
\mathrm{Tr}\rho^n_{\mathrm{red.}}
&= \left(\mathrm{det}\left(\mathbf{A}+4\mathbf{X}\right)\right)^{\frac{n}{2}} \left(\mathrm{det}\,\mathbf{M}^{(\tq=2)}_n\right)^{-\frac{1}{2}} 
\end{split}    
\end{align}
where $\mathbf{M}^{(\tq=2)}_n$ is a block-circulant matrix defined with the $\mathbf{M}^{(\tq=2)}_n=\{\mathbf{A}+2\mathbf{X},\mathbf{X},\mathbf{0},\cdots,\mathbf{0},\mathbf{X}\}$ blocks. For block-circulant matrices it is well known that the determinant is given by
\begin{align}
    \begin{split}
\mathrm{det}\left(\mathbf{M}^{(\tq=2)}_n\right)&=\prod_{k=1}^n \mathrm{det}\left(\mathbf{\Lambda}_k\right)
\\
\mathbf{\Lambda}_k&=\sum_{j=0}^{n-1}e^{\frac{2\pi i kj}{n}}\,{\mathbf{M}^{(\tq=2)}_n}_{k}\;.        
    \end{split}
\end{align}
Using this expression one can find a closed analytic form for $\mathrm{Tr}\rho^n_{\mathrm{red.}}$ in terms of $\mathbf{A}$ and $\mathbf{X}$. The final result coincides with Rényi entropies directly worked out from the covariance matrix method. This result not only reflects a well-known fact about Gaussian bipartite states where the Rényi-2 entropy is simply proportional to the determinant of the covariance matrix, but also shows that for generic $n$, Rényi-n entropy is a function of the determinants of blocks (and their inverses) of the covariance matrix. 

\begin{figure*}[t]
\begin{center}
\includegraphics[scale=0.3]{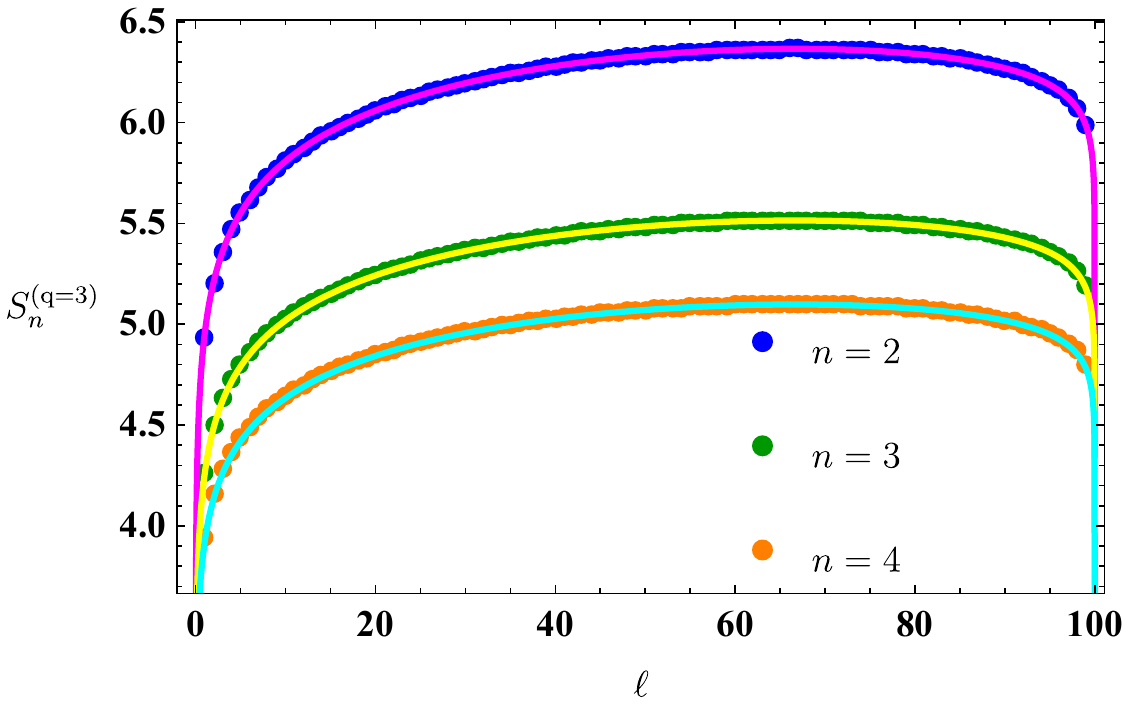}
\includegraphics[scale=0.3]{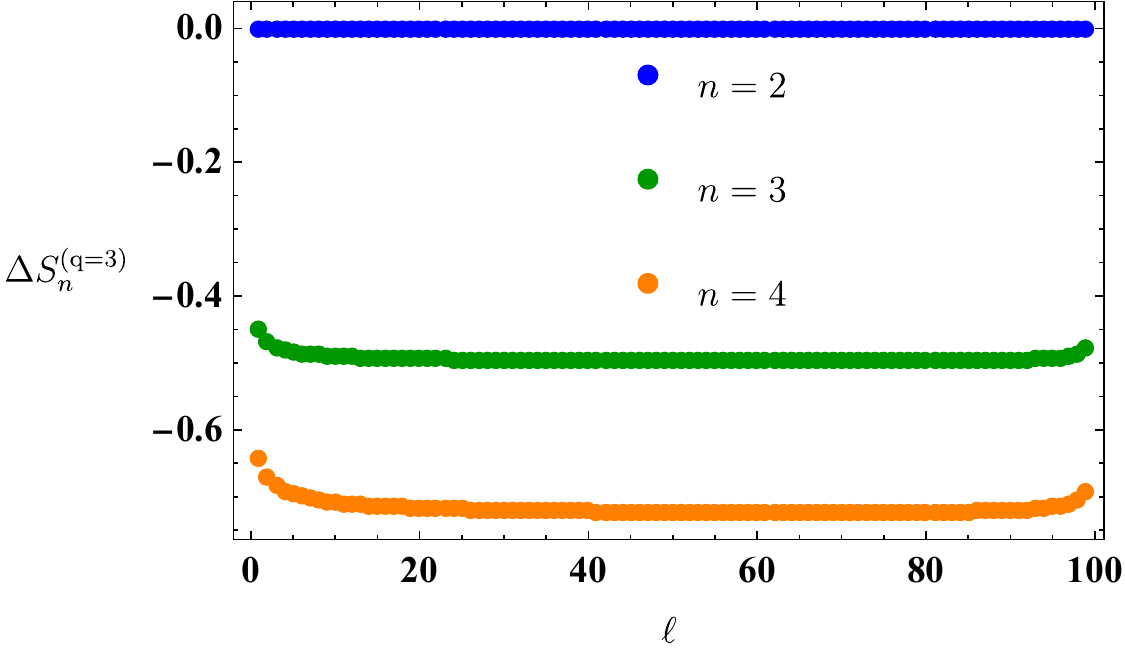}
\includegraphics[scale=0.3]{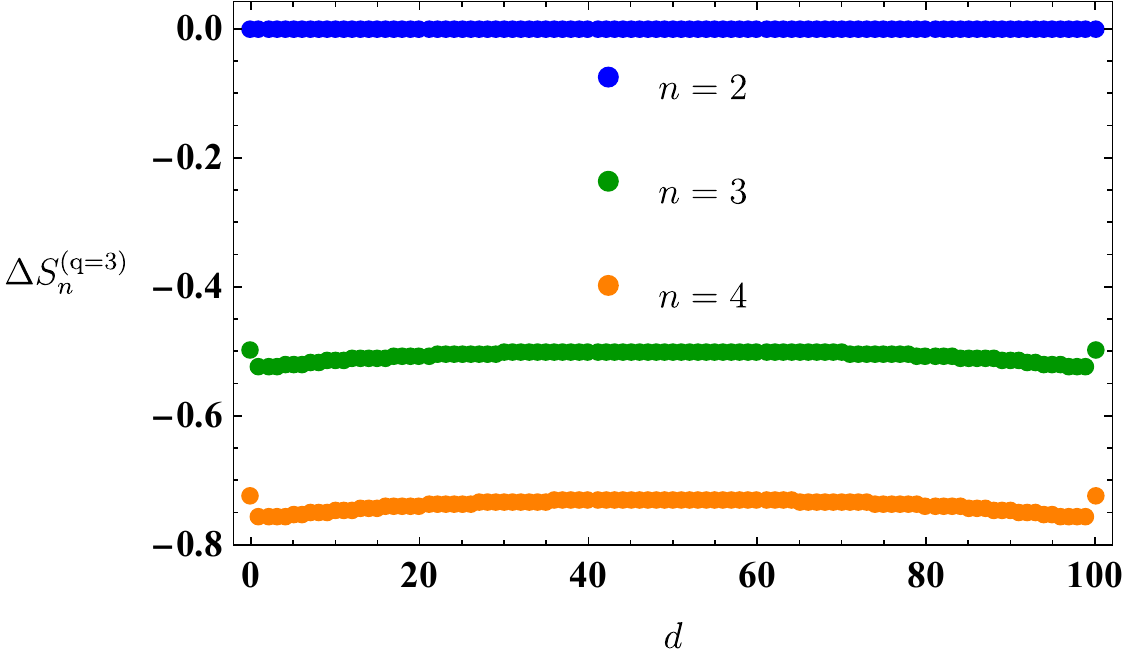}
\end{center}
\caption{
$\tq=3$ multi-entropy for scalar field theory.
\textit{Left:} Three adjacent intervals on a circle with circumference $L$. We set $\ell_A=\ell_B=\ell$ and $\ell_C=L-2\ell$. The data points are found from our numerical method where we set $m=10^{-5}$ and $L=200$. The (magenta, yellow, and cyan) solid curves corresponds to the compactified version of Eq. \eqref{eq:Snq3}, namely $S_n^{(3)}=\frac{1}{12}\left(1+\frac{1}{n}\right)\log\left(\epsilon^{-3}\sin\frac{\ell\pi}{L}\sin\frac{(L-\ell)\pi}{L}\sin\frac{(L-2\ell)\pi}{L}\right)+\kappa_n^{(3)}$ where we numerically find $\epsilon=3.669\times 10^{-8}$ such that the best fit leads to $\kappa_2^{(3)}=0$. With this choice we find $\kappa_3^{(3)}=-0.144$, and $\kappa_4^{(3)}=-0.207$. \textit{Middle:} The excess for the same configuration as the left panel. The excess for $n=0$ vanishes and takes negative values for $n>2$. \textit{Right:} here we present the excess of multi-entropy when the configuration is disjoint, namely $A$ and $B$ with $\ell_A=\ell_B=\ell$ are on a circle where their smaller distance is denoted by $d$. 
}
\label{fig:scalarNum1}
\end{figure*}

\subsection{Case $\tq=3$}
To work out the $\tq=3$ case, we must be careful about the structure of the reduced density matrix given in Eq. \eqref{eq:rhored}. For the case of $\tq=3$ the blocks in matrix $\mathcal{M}$ possess an internal structure due to two parties that we call $A$ and $B$. For instance the first block
$$\mathcal{M}_{11}=\begin{pmatrix}    \mathbf{A}_{AA} + 2\mathbf{X}_{AA} & \mathbf{A}_{AB} + 2\mathbf{X}_{AB} \\ \mathbf{A}_{BA} + 2\mathbf{X}_{BA} & \mathbf{A}_{BB}+2\mathbf{X}_{BB}\end{pmatrix}$$
and the same structure applies to other blocks as well.

Considering this structure we find the following for the $\tq=3$ case
\begin{align}\label{eq:Znq3result}
\begin{split}
\mathcal{E}&= 
\left(\mathrm{det}\left(\mathbf{A}+4\mathbf{X}\right)\right)^{\frac{n^2}{2}}
\left(\mathrm{det}\;\mathbf{M}^{(\tq=3)}_n\right)^{-\frac{1}{2}}
\end{split}    
\end{align}
where $\mathbf{M}^{(3)}_{n}$ is a $2n^2\times 2n^2$ block matrix that we find the block elements from appropriate contractions. For instance for $n=2$ the explicit form is given by
\begin{align}
\begin{split}
\mathbf{M}^{(\tq=3)}_2 &=
\begin{pmatrix}
\mathbf{\tilde{M}}_{1} & \mathbf{\tilde{M}}_{2} & \mathbf{\tilde{M}}_{3} & \mathbf{\tilde{M}}_{4}
\\
\mathbf{\tilde{M}}_{2} & \mathbf{\tilde{M}}_{1} & \mathbf{\tilde{M}}_{4} & \mathbf{\tilde{M}}_{3}
\\
\mathbf{\tilde{M}}_{3} & \mathbf{\tilde{M}}_{4} & \mathbf{\tilde{M}}_{1} & \mathbf{\tilde{M}}_{2}
\\
\mathbf{\tilde{M}}_{4} & \mathbf{\tilde{M}}_{3} & \mathbf{\tilde{M}}_{2} & \mathbf{\tilde{M}}_{1}
\end{pmatrix}  
\\
\mathbf{\tilde{M}}_{1}&=
\begin{pmatrix}    \mathbf{A}_{AA} + 2\mathbf{X}_{AA} & \frac{1}{2}\mathbf{A}_{AB} + \mathbf{X}_{AB} \\ \frac{1}{2}\mathbf{A}_{BA} + \mathbf{X}_{BA} & \mathbf{A}_{BB}+2\mathbf{X}_{BB} \end{pmatrix}
\\
\mathbf{\tilde{M}}_{2}&=
\begin{pmatrix}    2\mathbf{X}_{AA} & \mathbf{X}_{AB} \\  \mathbf{X}_{BA} & 0 \end{pmatrix}
\\
\mathbf{\tilde{M}}_{3}&=
\begin{pmatrix}    0 & \mathbf{X}_{AB} \\  \mathbf{X}_{BA} & 2\mathbf{X}_{BB} \end{pmatrix}
\\
\mathbf{\tilde{M}}_{4}&=
\begin{pmatrix}    0 & \frac{1}{2}\mathbf{A}_{AB} + \mathbf{X}_{AB} \\ \frac{1}{2}\mathbf{A}_{BA} + \mathbf{X}_{BA} & 0 \end{pmatrix}
\end{split}    
\end{align}


Although the final expression for $\tq=3$ case given in Eq. \eqref{eq:Znq3result} is easy to calculate numerically, a very interesting open question is to find a closed form for it in terms $n$ such that one can analytically continue the result to find the $n\to1$ multi-entropy.

Plugging Eq. \eqref{eq:Znq3result} into Eq. \eqref{eq:MEq3} given the multi-entropy. In Fig. \ref{fig:scalarNum1} in the left panel we show numerical calculation of $S_n^{(3)}$ using Eq. \eqref{eq:Znq3result} for the case of adjacent intervals and $n=2,3,4$. One can see perfect agreement with the corresponding CFT result for $c=1$. In the middle panel we show the excess of multi-entropy for this case where we find vanishing excess for $n=2$ case in agreement with our CFT result, Eq. \eqref{eq:deltaSn2Scalar}. For $n>2$ we find non-constant negative values for the excess where the deviation from a constant is increasing with $n$. An extrapolation to $n=1$ would lead to a \textit{constant positive} value for $\Delta S^{(3)}$. In the right panel a similar behavior is found for disjoint $A$ and $B$. We observe that for larger $\ell$ the excess for $n>2$ gets less sensitive to $d$, indicating that this non-constant behavior is partially a lattice effect. Let us also emphasize that the same non-constant behavior is also observed on an infinite lattice, where the finite size effect is suppressed. 

\begin{figure*}[t!]
\begin{center}
\includegraphics[scale=0.3]{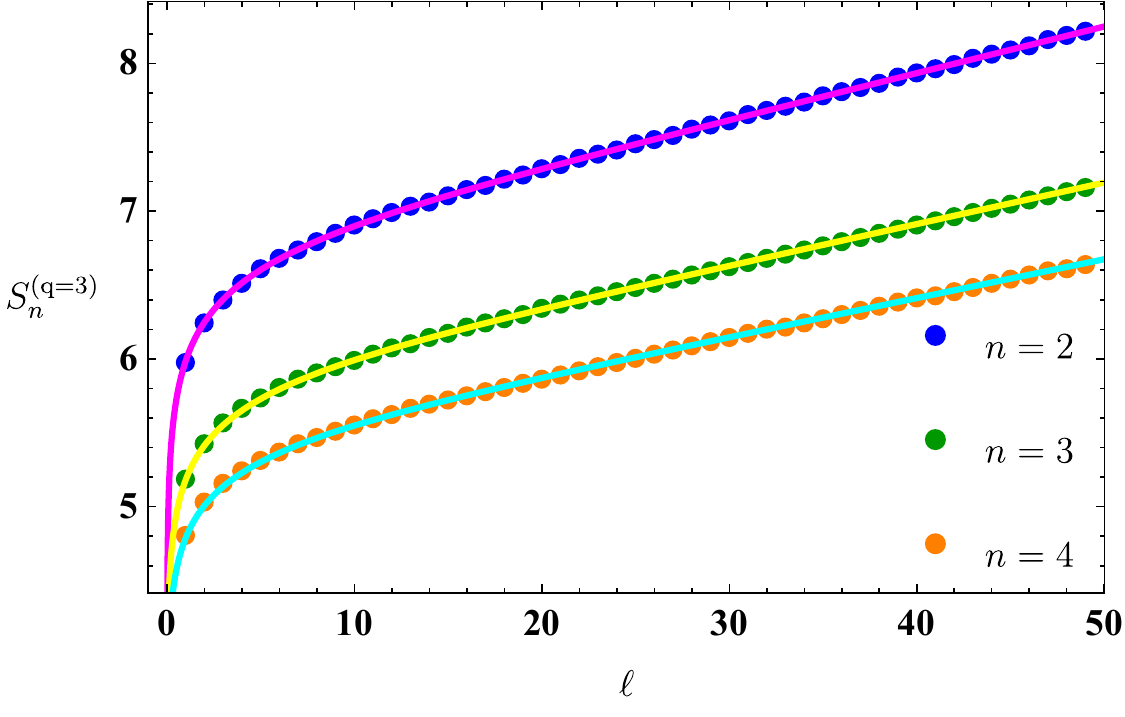}
\hspace{4mm}
\includegraphics[scale=0.3]{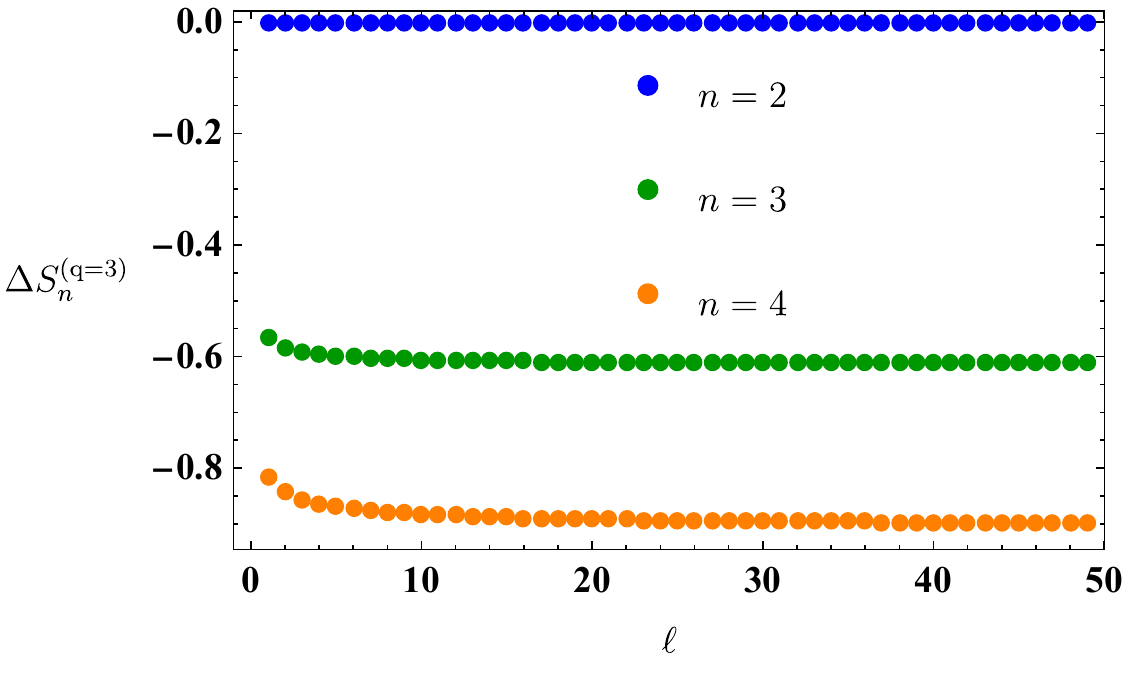}
\end{center}
\caption{
$\tq=3$ multi-entropy at finite temperature for free scalar field theory.
\textit{Left:} Three adjacent intervals on infinite lattice. We set $\ell_A=\ell_B=\ell$ and the complement as $\ell_C$. We set $m=10^{-5}$ and $\beta=1/T=50$. \textit{Left:} multi-entropy for $n=2,3,4$ as the temperature increases, multi-entropy approaches a volume law scaling. The (magenta, yellow, and cyan) solid curves corresponds to the finite temperature version of Eq. \eqref{eq:Snq3}, namely $S_n^{(3)}=\frac{1}{12}\left(1+\frac{1}{n}\right)\log\left(\epsilon^{-3}\sinh^2\frac{\ell\pi}{\beta}\sinh\frac{2\ell\pi}{\beta}\right)+\kappa_n^{(3)}$ where we numerically find $\epsilon=9.209\times 10^{-9}$ such that the best fit leads to $\kappa_2^{(3)}=0$. With this choice we find $\kappa_3^{(3)}=-0.140$, and $\kappa_4^{(3)}=-0.200$. \textit{Right:} the excess of multi-entropy for the same cases shown in the left panel. For $n=2$ for any value of $\beta$ we find the excess to vanish and for larger $n$ the excess is always negative. }
\label{fig:scalarNum2}
\end{figure*}

Since we are dealing with bosonic Gaussian states and the finite temperature state of quadratic bosonic Hamiltonians, including the free scalar theory is also a Gaussian state, our numerical procedure can be applied to this state as well. We have shown the behavior of multi-entropy and its excess in this case in Fig. \ref{fig:scalarNum2}. From the numerical results one can observe that the behavior of multi-entropy approaches a volume law as we put the system at finite temperature. As we have shown for a choice of $\beta=50$ in this plot, the excess vanishes for $n=2$ case. This behavior is observed for various values of $\beta$. For $n>2$, as have been shown in the Fig. for $n=3$, we always find a negative value for the excess.

\subsection*{Dihedral Measure}
To calculate the dihedral measure for the free scalar theory, we follow a similar procedure we described for multi-entropy. Applying the appropriate contractions for $n=2$ leads to 
\be
\mathbf{D}_{2n}=
\begin{pmatrix}
\mathbf{\tilde{M}}_{1} & \mathbf{\tilde{M}}_{2} & \mathbf{\tilde{M}}_{4} & \mathbf{\tilde{M}}_{3}
\\
\mathbf{\tilde{M}}_{2} & \mathbf{\tilde{M}}_{1} & \mathbf{\tilde{M}}_{3} & \mathbf{\tilde{M}}_{4}
\\
\mathbf{\tilde{M}}_{4} & \mathbf{\tilde{M}}_{3} & \mathbf{\tilde{M}}_{1} & \mathbf{\tilde{M}}_{2}
\\
\mathbf{\tilde{M}}_{3} & \mathbf{\tilde{M}}_{4} & \mathbf{\tilde{M}}_{2} & \mathbf{\tilde{M}}_{1}
\end{pmatrix}  
\ee
where
\be
\mathcal{D}_{2n}=\frac{1}{1-n}\frac{1}{2n}\log\left(\left(\mathrm{det}\left(\mathbf{A}+4\mathbf{X}\right)\right)^{\frac{2n}{2}}
\left(\mathrm{det}\;\mathbf{D}_{2n}\right)^{-\frac{1}{2}}\right)
\ee

\begin{figure*}[t!]
\begin{center}
\includegraphics[scale=0.3]{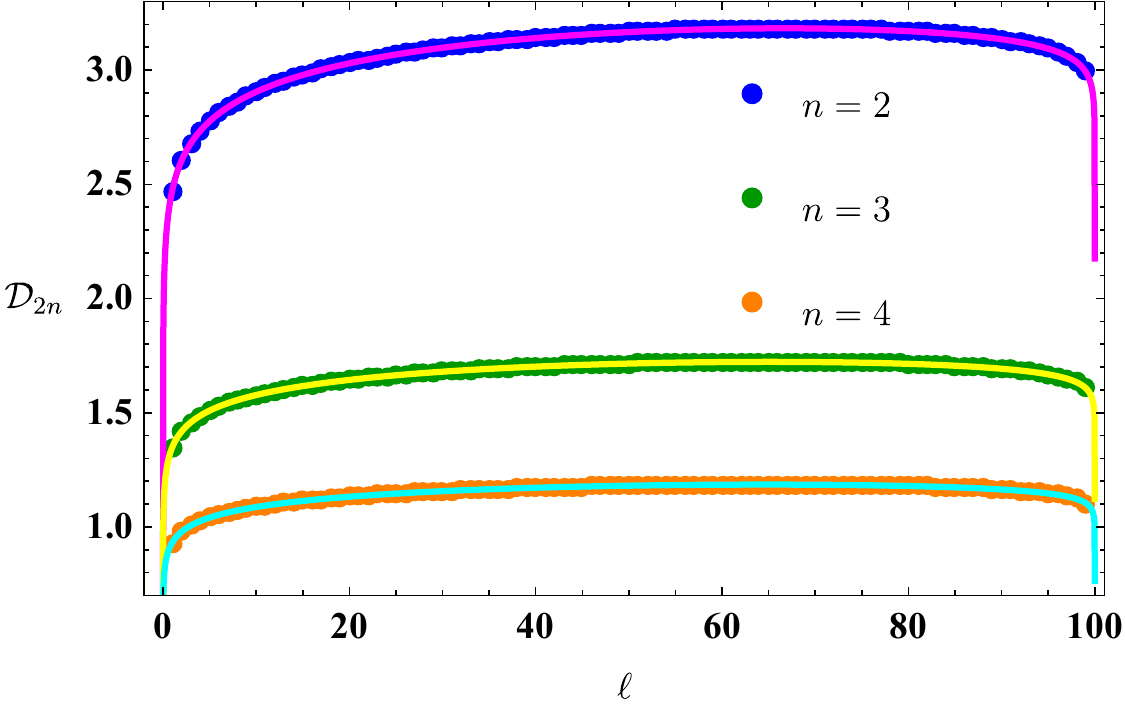}
\hspace{4mm}
\includegraphics[scale=0.3]{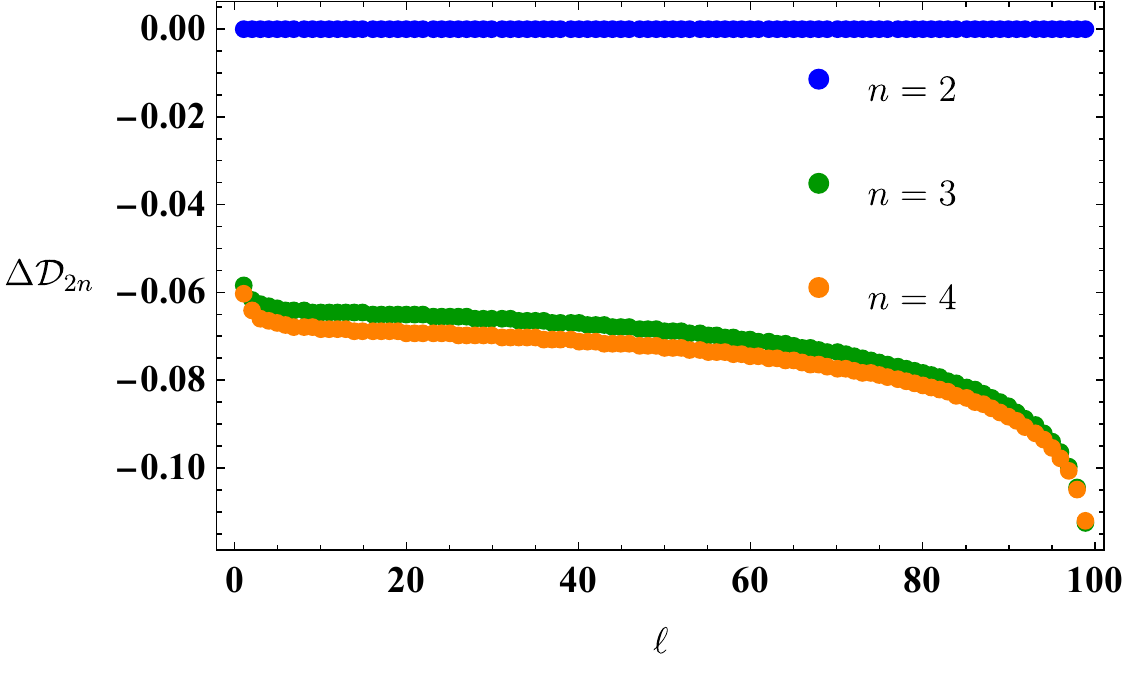}
\end{center}
\caption{Dihedral measure for free scalar theory. We have considered three adjacent intervals with $\ell_A=\ell_B=\ell$ and $\ell_C=L-2\ell$ with $L=200$ and $m=10^{-5}$. \textit{Left:} The data points correspond to our numerical results, and the solid curves correspond to the analytical result 
$\mathcal{D}_{2n}=\frac{1}{1-n}\frac{1}{2n}\left(\frac{1}{6}\frac{n^2-1}{n} \log(\frac{1}{\epsilon}\sin\frac{\ell\pi}{L})+\frac{1}{6}\frac{n^2-1}{n} \log(\frac{1}{\epsilon}\sin\frac{(L-2\ell)\pi}{L}) +\frac{1}{12}\frac{n^2+2}{n} \log(\frac{1}{\epsilon}\sin\frac{(\ell-L)\pi}{L})\right)+\Delta \mathcal{D}_{2n}$
and we find $\epsilon=3.675\times10^{-8}$ such that the best fit gives $\Delta \mathcal{D}_{4}=0$ and correspondingly we find $\Delta \mathcal{D}_6=0.0329$ and $\Delta \mathcal{D}_8=0.0359$. \textit{Right:} The excess for the same configuration shown in the left panel.}
\label{fig:scalarNumD}
\end{figure*}

For $n>2$ one can find $\mathbf{D}_{2n}$ similarly by applying the appropriate contractions. In Fig. \ref{fig:scalarNumD} we have shown our numerical simulation results and compared the results with the CFT exact results.

\clearpage

\section{Numerical multi-entropy for transverse-field Ising Model}
In this section, we perform a numerical study of the multi-entropy for the transverse-field Ising chain, one of the paradigmatic examples of a quantum phase transition.
We first focus on the case of adjacent intervals and show that the multi-entropy excess quantitatively reproduces the CFT prediction at criticality.
A finite-size scaling analysis of the adjacent multi-entropy excess allows us to extract the critical field and the central charge with high precision.
We then turn to the disjoint configuration, demonstrating that the separation dependence of the multi-entropy excess is consistent with the CFT prediction and in agreement with exact results derived via the Jordan-Wigner transformation.

\subsection{Model and setup}
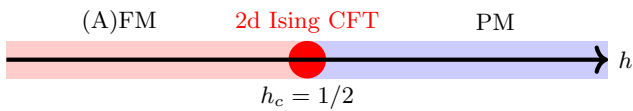
\begin{figure}[tbp]
    \centering
    \begin{tikzpicture}
        \fill[fill=red!20 ] (-4, -0.25) rectangle (0, 0.25);
        \fill[fill=blue!20] ( 0, -0.25) rectangle (4, 0.25);

        \fill[red] (0, 0) circle [radius=0.25];

        \node at (-2.5,0.5) {(A)FM};
        \node[red] at (0,0.5) {2d Ising CFT};
        \node at ( 2.5,0.5) {PM};

        \node at (0,-0.5) {$h_c = 1/2$};

        \draw[ultra thick, ->] (-4, 0) -- (4, 0) node[right] {$h$};
    \end{tikzpicture}
    \caption{Ground state phase diagram of the transverse-field Ising model.}
    \label{fig:phase_diagram}
\end{figure}

We study the one-dimensional transverse-field Ising model with Hamiltonian
\begin{equation}
    H = - J \sum_i S^{(z)}_i S^{(z)}_{i+1} - B_x \sum_i S^{(x)}_i,
\end{equation}
where $S^{(\alpha)}_i$ are spin-$1/2$ operators in $\alpha$-direction on site $i$.
The coupling $J$ determines the nearest-neighbour interaction, while $B_x$ is a homogeneous transverse magnetic field.
The interaction is ferromagnetic (FM) for $J>0$ and antiferromagnetic (AFM) for $J<0$.
Introducing the ratio $h \coloneqq |B_x/J|$, the ground-state phase diagram is shown in Fig.~\ref{fig:phase_diagram}.
A quantum critical point at $h_c \coloneqq 1/2$ separates the (A)FM phase ($h < h_c$) from the paramagnetic (PM) phase ($h > h_c$).
The low-energy physics at the critical point is effectively described by the two-dimensional Ising CFT with central charge $c=1/2$.

\begin{figure*}[tbp]
    \centering
    \begin{tikzpicture}
        \coordinate (L) at (-5.5,0);
        \foreach \x [count=\n] in {0, 3, 6}
            \draw[very thick,fill=black] (\x+1,0) circle(0.2) (\x-1,0) circle(0.2) (-\x+1,0) circle(0.2) (-\x-1,0) circle(0.2) node at (\x,0) {$\cdots$} node at (-\x,0) {$\cdots$};

        \draw[very thick,rounded corners=5pt,orange] (-6-1.4,-0.4) rectangle (-6+1.4, 0.4) node[midway,above=15pt] {$C$};
        \draw[very thick,rounded corners=5pt,red]    (-3-1.4,-0.4) rectangle (-3+1.4, 0.4) node[midway,above=15pt] {$A$};
        \draw[very thick,rounded corners=5pt,orange] ( 0-1.4,-0.4) rectangle ( 0+1.4, 0.4) node[midway,above=15pt] {$C$};
        \draw[very thick,rounded corners=5pt,blue]   ( 3-1.4,-0.4) rectangle ( 3+1.4, 0.4) node[midway,above=15pt] {$B$};
        \draw[very thick,rounded corners=5pt,orange] ( 6-1.4,-0.4) rectangle ( 6+1.4, 0.4) node[midway,above=15pt] {$C$};

        \draw[decorate,decoration={brace,mirror,amplitude=5mm}] (-6-1.3,-0.7) -- (-6+1.3,-0.7) node[midway,below=15pt] {$(L - \ell_A - \ell_B - \ell)/2$};
        \draw[decorate,decoration={brace,mirror,amplitude=5mm}] (-3-1.3,-0.7) -- (-3+1.3,-0.7) node[midway,below=15pt] {$\ell_A$};
        \draw[decorate,decoration={brace,mirror,amplitude=5mm}] ( 0-1.3,-0.7) -- ( 0+1.3,-0.7) node[midway,below=15pt] {$\ell$};
        \draw[decorate,decoration={brace,mirror,amplitude=5mm}] ( 3-1.3,-0.7) -- ( 3+1.3,-0.7) node[midway,below=15pt] {$\ell_B$};
        \draw[decorate,decoration={brace,mirror,amplitude=5mm}] ( 6-1.3,-0.7) -- ( 6+1.3,-0.7) node[midway,below=15pt] {$(L - \ell_A - \ell_B - \ell)/2$};
    \end{tikzpicture}
    \caption{Partition of the spin chain into subsystems $A$, $B$, and $C$ used for adjacent ($\ell=0$) and disjoint ($\ell>0$) setups.}
    \label{fig:spinchain_setup}
\end{figure*}
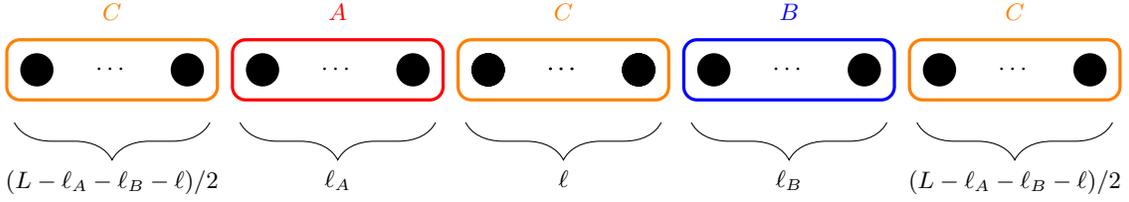

Henceforth, we restrict to the ferromagnetic case and set $J=1$.
Calculations are performed on chains of length $L$ with open boundary conditions (OBC).
As sketched in Fig.~\ref{fig:spinchain_setup}, the open chain is partitioned into three subsystems: two finite blocks $A$ and $B$ of length $\ell_A$ and $\ell_B$, separated by a intermediate buffer of length $\ell$, and the complement $C$ at the edges of the chain and the buffer.
The adjacent setup corresponds to $\ell=0$, while the disjoint setup corresponds to $\ell>0$.

\subsection{Exact calculation by two-site density matrix}\label{sec:exact}
Before turning to numerical simulations, we analyze setups with $\ell_A = \ell_B = 1$ in the thermodynamic limit $L \to \infty$.
In this case, the two-site reduced density matrix $\rho_{AB}$ of the ground state can be computed exactly, following the method of~\cite{Osborne:2002zz}.

The reduced density matrix for a single site~$i$ is expanded as
\begin{equation}
    \rho_i = \frac{1}{2} \sum_\alpha \langle \sigma_i^{(\alpha)} \rangle \sigma_i^{(\alpha)},
\end{equation}
while the two-site reduced density matrix for site~$i$ and site~$j$ is given by
\begin{equation}
    \rho_{i,j} = \frac{1}{4} \sum_{\alpha, \beta} \langle \sigma_i^{(\alpha)} \sigma_j^{(\beta)} \rangle \sigma_i^{(\alpha)} \sigma_j^{(\beta)},
\end{equation}
where $\sigma_i^{(\alpha)}$ are the Pauli matrices at site~$i$ with the convention $\sigma_i^{(0)} = \openone_i$.

The exact solution of the transverse-field Ising chain, or more generally the one-dimensional XY model, is obtained in~\cite{Lieb:1961fr,Barouch:1970ryz} by mapping the spin operators to fermionic operators via the Jordan--Wigner transformation~\cite{Jordan:1928wi}.
Non-trivial expectation values for the ground state at the critical point are calculated in~\cite{Pfeuty:1970qrn} as
\begin{align}
    \langle \sigma_i^{(x)} \rangle &= \frac{2}{\pi}, \\
    \langle \sigma_i^{(x)} \sigma_{i+r}^{(x)} \rangle &= \frac{4}{\pi^2} \frac{1}{4r^2-1} + \langle \sigma_i^{(x)} \rangle^2, \\
    \langle \sigma_i^{(z)} \sigma_{i+r}^{(z)} \rangle &= \left( \frac{4^r}{2\pi} \right)^r \frac{{H(r)}^4}{H(2r)}, \\
    \langle \sigma_i^{(y)} \sigma_{i+r}^{(y)} \rangle &= - \frac{1}{4r^2-1} \langle \sigma_i^{(z)} \sigma_{i+r}^{(z)} \rangle,
\end{align}
where $H(r)$ is defined as
\begin{equation}
    H(r) \coloneqq \prod_{k=1}^{r-1} k^{r-k}
\end{equation}
and $r = \ell+1$ is the distance between the two sites.

The Rényi entropy and Rényi multi-entropy are then computed from the two-site reduced density matrix $\rho_{i, i+r}$ as
\begin{widetext}
\begin{align}
    S_2^{(3)}     & = - \frac{1}{2} \log\left[ \left\{ \frac{1}{2} \left( \frac{4^r}{2\pi} \right)^r \frac{H(r)^4}{H(2r)} \right\}^4 \left( 1 + \frac{1}{(4r^2-1)^4} \right) + \frac{2^{12}}{\pi^8}\frac{r^8}{(4r^2-1)^4} + \frac{1}{\pi^2} \sum_{n=0, 1, 2} \left( \frac{16r^2}{\pi^2(4r^2-1)} \right)^n + \frac{\pi^4+32}{16\pi^4} \right], \\
    S_2^{(2)}(AB) & = - \log\left[ \left\{ \frac{1}{2} \left( \frac{4^r}{2\pi} \right)^r \frac{H(r)^4}{H(2r)} \right\}^2 \left( 1 + \frac{1}{(4r^2-1)^2} \right) +\frac{64}{\pi^4} \frac{r^4}{(4r^2-1)^2} + \frac{\pi^2+8}{4\pi^2} \right], \\
    S_2^{(2)}(A)  & = S_2^{(2)}(B) = \log\left[ \frac{2\pi^2}{\pi^2+4} \right].
\end{align}
\end{widetext}
These exact analytical results are shown in Fig.~\ref{fig:Ising_disconnected_rdep_lattice}, where comparison is made with the numerical data obtained from the tensor network calculations described in the following section.

In the distant limit $r \to \infty$, the entropic quantities converge to the following values:
\begin{equation}
    \lim_{r \to \infty} S_2^{(3)} = \lim_{r \to \infty} S_2^{(2)}(AB) = 2S_2^{(2)}(A) = 2\log\left[ \frac{2\pi^2}{\pi^2+4} \right],\label{eq:exact_distantlimit}
\end{equation}
leading to vanishing multi-entropy excess $\kappa_2^{(3)}$ in this limit.
The leading asymptotic behavior of the multi-entropy excess is calculated as
\begin{equation}
    \kappa_2^{(3)} \sim \frac{e\pi^4}{2^{5/6} (\pi^2+4)^2 A^6} \frac{1}{\sqrt{r}} \quad (r \to \infty),
\end{equation}
where $A=1.282\cdots$ is the Glaisher--Kinkelin constant.
Although this result is for two-site entropies, it is noteworthy that this decay exponent of the multi-entropy excess coincides with that of the CFT calculation \eqref{eq:IsingCFT_excess_distant}.

\subsection{Tensor network calculation}
We now turn to numerical calculations based on the matrix product state representation of the ground state.
All tensor network calculations are performed with the Julia library \texttt{ITensor}~\cite{Fishman:2020gel,Fishman:2022sdc}.

\subsubsection{Method}
\begin{figure}[tbp]
    \centering
    \includegraphics[width=0.45\textwidth]{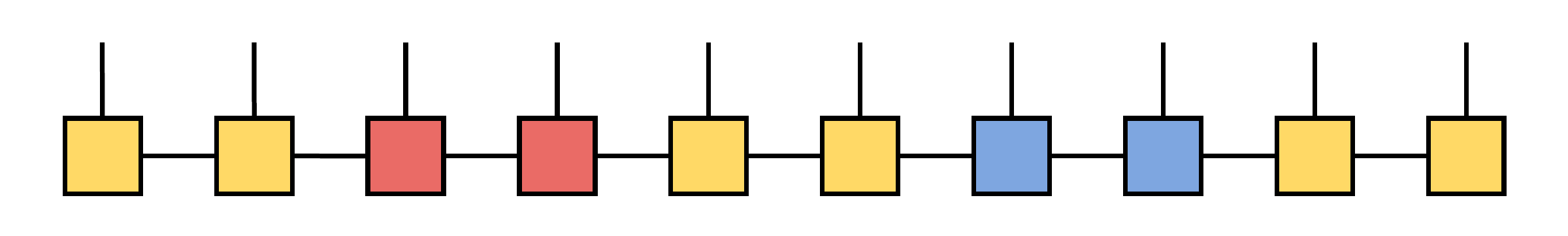}
    \includegraphics[width=0.45\textwidth]{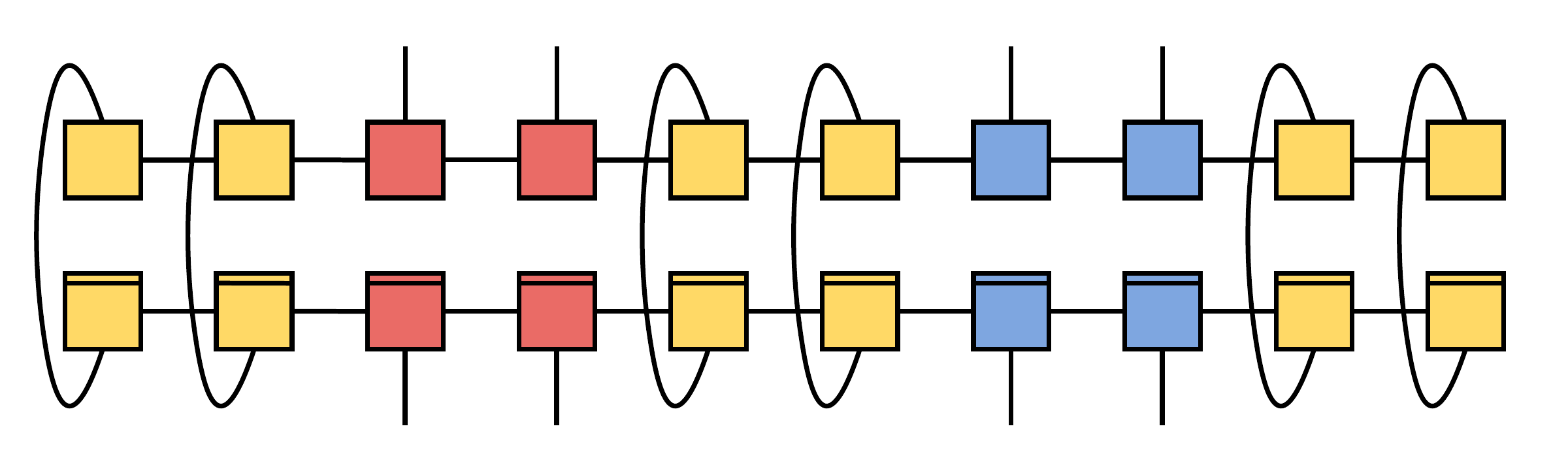}
    \includegraphics[width=0.45\textwidth]{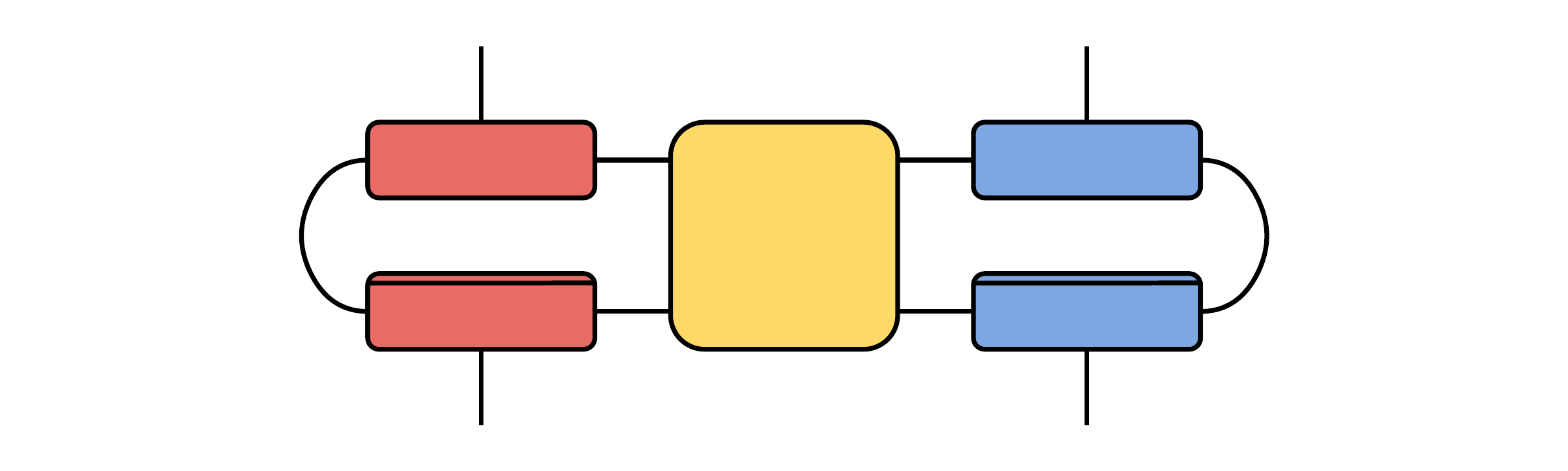}
    \caption{
        Schematic illustration of the construction of the reduced density matrix.
        Red, blue, and yellow tensors are related to the subsystems $A$, $B$, and $C$, respectively.
        \textit{Top:}~MPS representation of the ground state of the transverse-field Ising chain.
        \textit{Middle:}~Naive construction of the reduced density matrix $\rho_{AB}$ for adjacent intervals $A$ and $B$.
        \textit{Bottom:}~Construction of the truncated reduced density matrix used in the numerical calculation.
    }
    \label{fig:rdm}
\end{figure}

The matrix product state (MPS) representation of the ground state (top panel of Fig.~\ref{fig:rdm}) is obtained via the density matrix renormalisation group (DMRG) algorithm~\cite{White:1992zz}.
For two intervals $A$ and $B$, one could in principle build the reduced density matrix by contracting every physical (site) index outside $A \cup B$ in the MPS with its conjugate (middle panel of Fig.~\ref{fig:rdm}).
However, we adopt a more efficient alternative method used in~\cite{Ruggiero:2016aqr}, in which transfer matrices play a crucial role.

First, we set the orthogonality center of the MPS at the right edge of the subsystem $A$: tensors in $A$ are therefore left-orthogonal, whereas those in $B$ (and the intermediate buffer, if any) are right-orthogonal.
With this choice, the trace over the region to the left of $A$ comes down to a single contraction of link indices at the left edge of $A$; the same applies to the right edge of $B$.

\begin{figure}[tbp]
    \centering
    \includegraphics[width=0.45\textwidth]{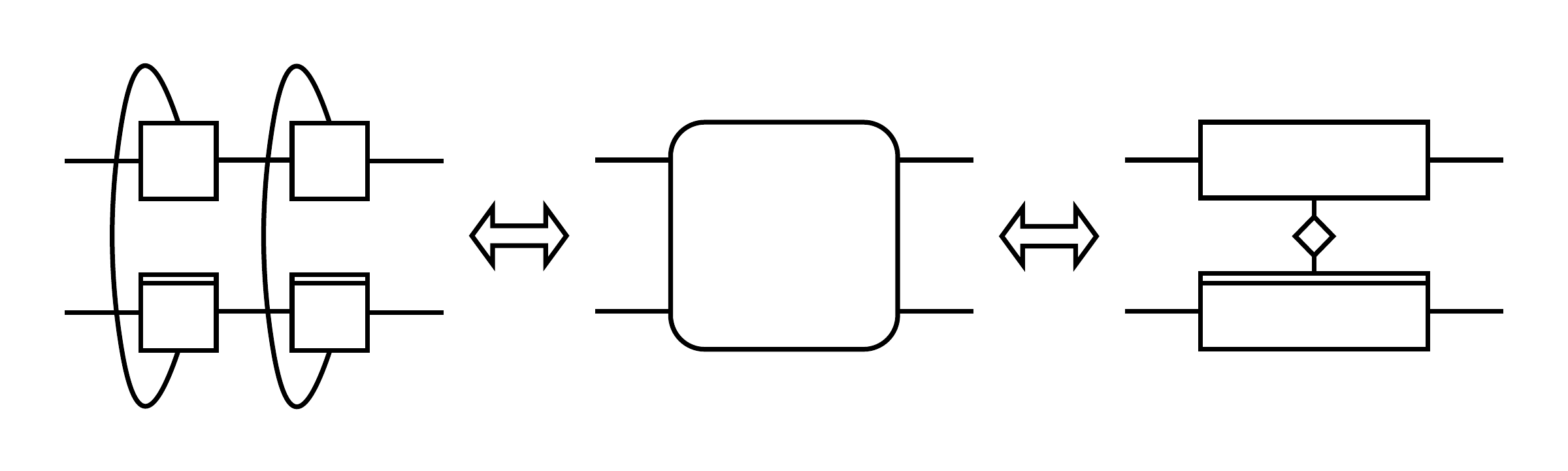}
    \includegraphics[width=0.45\textwidth]{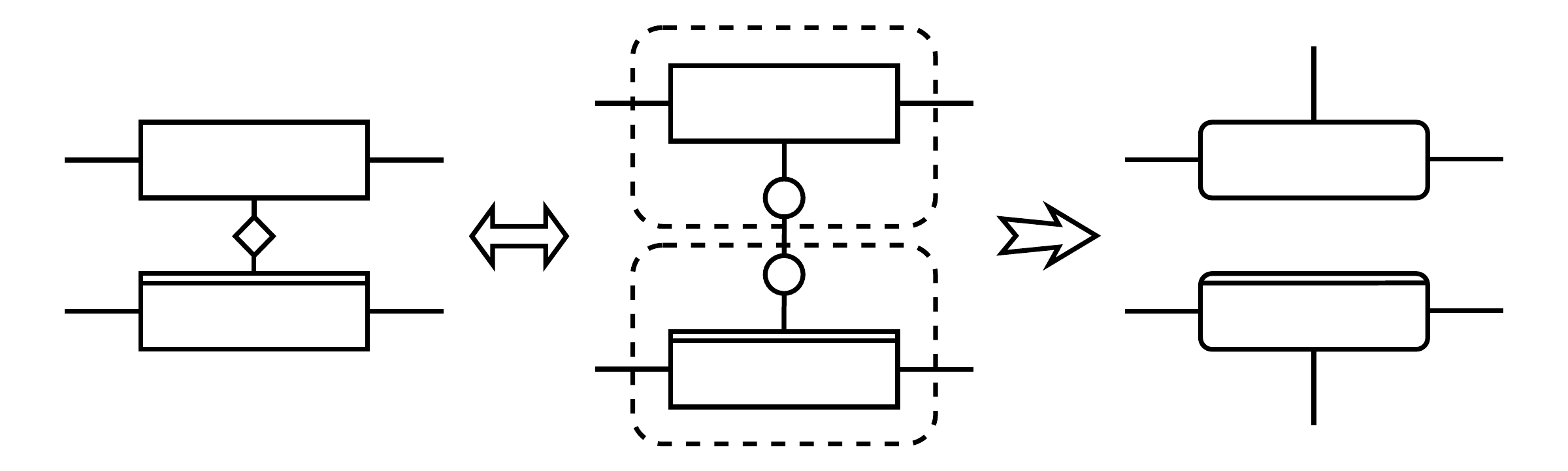}
    \caption{
        Schematic illustration of the construction of the transfer matrix and the truncated bra and ket tensors.
        \textit{Top:}~Transfer matrix (rounded square-shaped tensor) is built by contracting physical indices of the tensors in the subsystem.
        One can then perform a singular value decomposition on the transfer matrix.
        The rhombus-shaped tensor represents the diagonal singular value matrix.
        \textit{Bottom:}~The square roots of the diagonal singular value matrix, represented by the circle-shaped tensor, are absorbed into the bra and ket tensors, forming the truncated bra and ket tensors (rounded rectangle-shaped tensors).
    }
    \label{fig:transf}
\end{figure}

Next, we contract every physical index with its conjugate within each block $A$, $B$, and the intermediate buffer.
This operation leaves a tensor whose remaining indices are only the link indices that connect the block to its surroundings; this tensor constitutes the transfer matrix of the block (top panel of Fig.~\ref{fig:transf}).

Finally, we reconstruct a truncated reduced density matrix by factorising the transfer matrices of $A$ and $B$.
We perform a singular value decomposition (SVD) on each transfer matrix with respect to the bra and ket indices, and retain the largest $\chi_{\text{RDM}}$ singular values (top panel of Fig.~\ref{fig:transf}).
The square roots of the diagonal singular value matrix are absorbed into the bra and ket tensors, producing truncated tensors whose effective Hilbert space dimension is $\chi_{\text{RDM}}$.
Contracting the truncated bra and ket tensors for subsystems $A$ and $B$ with the transfer matrix for the intermediate buffer, we obtain the truncated reduced density matrix for the disjoint setup (bottom panel of Fig.~\ref{fig:rdm}).

In the DMRG calculations, we chose the maximum bond dimension $\chi_{\text{DMRG}} = 100$ and the truncation error $\epsilon_{\text{DMRG}} = 10^{-12}$.
The truncation of the transfer matrix is performed with $\chi_{\text{RDM}} = 32$.
These values were verified to yield converged results by comparison with calculations performed using other parameters, and were used consistently in all subsequent computations.

\subsubsection{Adjacent setup}
We first consider the adjacent setup ($\ell=0$) with equal block lengths ($\ell_A = \ell_B$).
We fix the ratio of the subsystem sizes as $\ell_A = \ell_B = L/4$ and vary the total system size $L$ in multiples of four, along with the transverse field $h$.
We will note the adjacent multi-entropy excess $\kappa_2^{(3)}$ in this setup as $\kappa(L, h)$.

\begin{figure*}[tbp]
    \centering
    \includegraphics[width=0.45\textwidth]{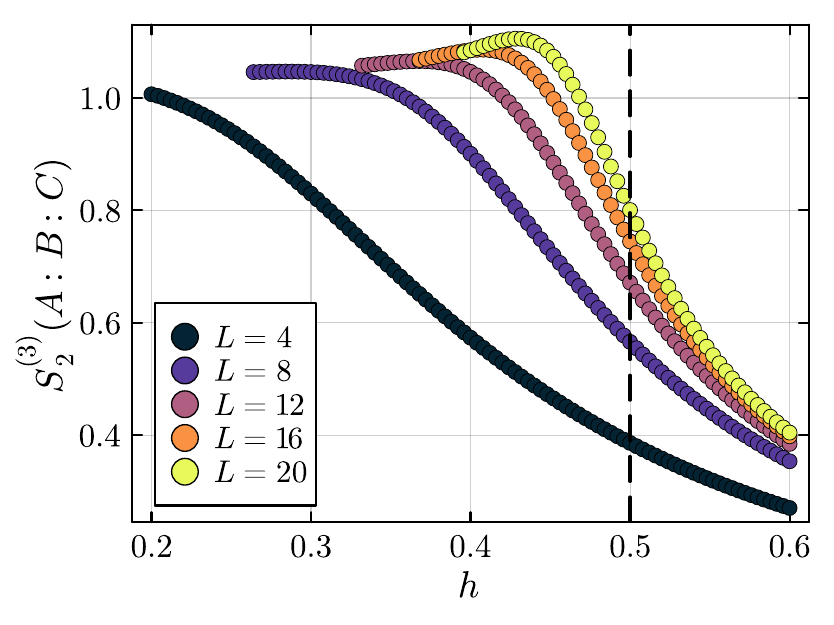}
    \includegraphics[width=0.45\textwidth]{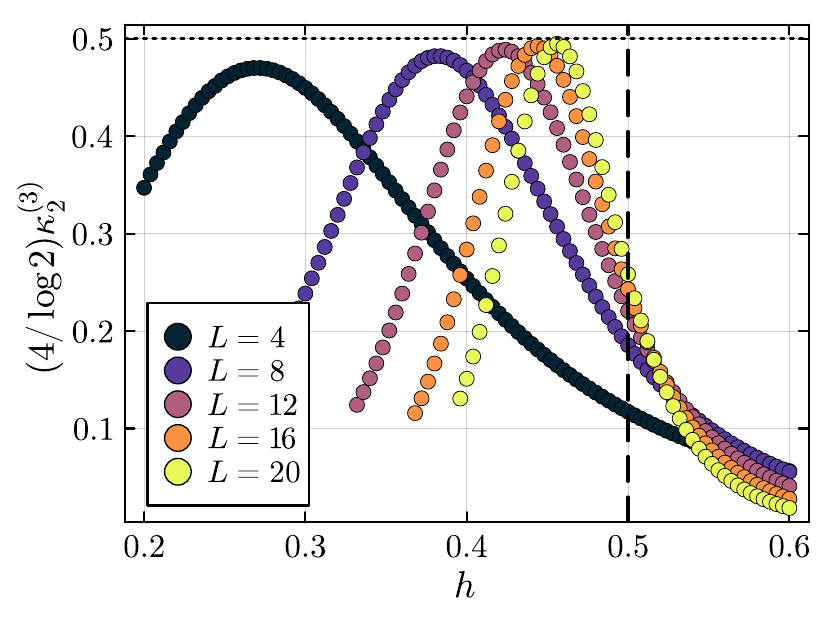}
    \includegraphics[width=0.45\textwidth]{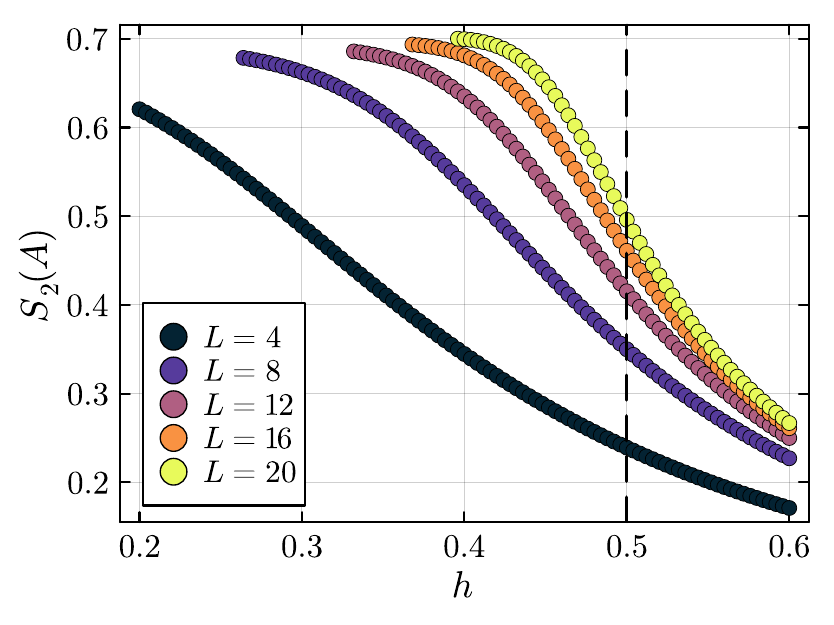}
    \includegraphics[width=0.45\textwidth]{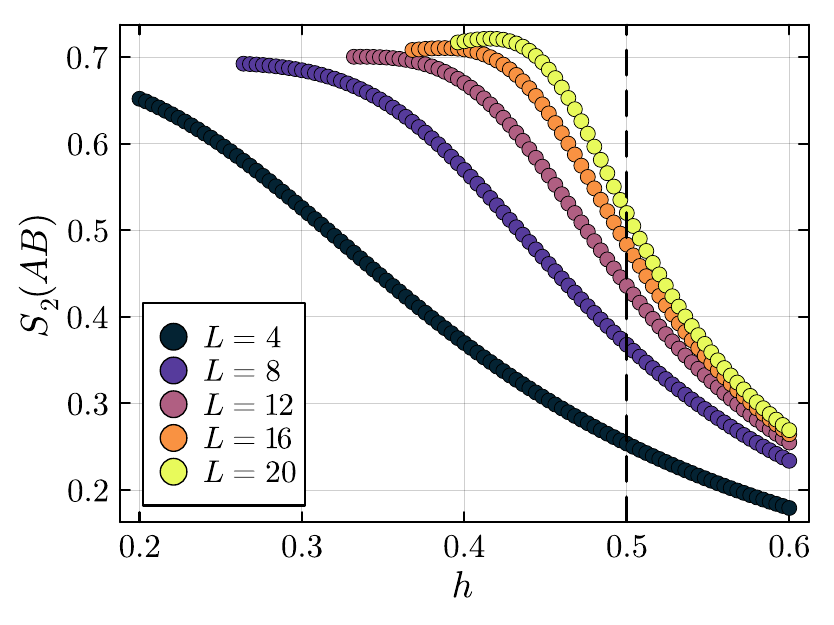}
    \caption{
        Entropies for the adjacent setup with $\ell_A = \ell_B = L/4$ with varying system size $L$ and transverse field $h$.
        Vertical dashed lines indicate the theoretical critical point $h_c = 1/2$.
        \textit{Top left:}~Multi-entropy $S_2^{(3)}$.
        \textit{Top right:}~Rescaled adjacent multi-entropy excess $(4/\log{2})\kappa_2^{(3)}$.
        The horizontal dotted line indicates the CFT prediction $c=1/2$.
        \textit{Bottom left:}~Single-interval Rényi entropy $S_2^{(2)}(A)$.
        \textit{Bottom right:}~Two-interval Rényi entropy $S_2^{(2)}(AB)$.
    }
    \label{fig:Ising_adjacent}
\end{figure*}

The computed entropic quantities are presented in Fig.~\ref{fig:Ising_adjacent}.
As the system size $L$ increases, the adjacent multi-entropy excess $\kappa(L, h)$ develops an increasingly sharp peak, whose height saturates from below toward a limiting value $\kappa_c$.
Each peak location defines a finite-size effective critical point, known as the \emph{pseudo-critical point} $h_{\text{pc}}(L)$, which lies below the true critical point $h_c$ but is expected to converge to it in the thermodynamic limit $L \to \infty$.
Notably, even for relatively small system sizes, the deviation of the peak value $\kappa_{\text{pc}}(L)$ from its limiting value $\kappa_c$ remains within $8\%$.

The pseudo-critical point $h_{\text{pc}}(L)$, and the peak value $\kappa_{\text{pc}}(L)$ exhibit finite-size scaling behavior
\begin{subequations}
    \label{eq:fss}
    \begin{align}
        h_c - h_{\text{pc}}(L) &\sim L^{-\alpha_h}, \label{eq:fss_h}\\
        \kappa_c - \kappa_{\text{pc}}(L) &\sim L^{-\alpha_\kappa}. \label{eq:fss_kappa}
    \end{align}
\end{subequations}
Extrapolating to the thermodynamic limit $L \to \infty$ by data points with $L \geq 8$, we obtain the true critical point $h_c$ and the limiting value $\kappa_c$ as summarized in Table~\ref{tab:adjacent_scaling}.
The details of the fitting procedure are described in~\ref{ssec:fit_adjacent}.
The true critical point $h_c$ obtained from the fitting procedure almost coincides with the theoretical value $h_c = 1/2$.
The limiting value $\kappa_c$ of the adjacent multi-entropy excess also matches the CFT prediction with the coincident limit prescription
\begin{equation}
    \kappa_2^{(3)} = \frac{c}{4} \log 2, \label{eq:adjacent_excess}
\end{equation}
where $c=1/2$ is the central charge of the Ising CFT.

The horizontal shift exponent $\alpha_h$ matches the naive finite-size scaling expectation $\alpha_h = 1/\nu$, where $\nu = 1$ is the correlation length exponent for the Ising universality class.
For standard entanglement or Rényi entropies (i.e., when $\tq = 2$), the vertical shift exponent arises from relevant operators associated with conical singularities and typically takes the form $s\varepsilon / n$, where $s$ is the number of entangling surfaces, $\varepsilon$ is the RG dimension of the leading conical operator (often dominated by the energy operator~\cite{Xavier:2011np}), and $n$ is the Rényi index~\cite{Cardy:2010zs}.
Extending the calculations to larger subsystem sizes and different configurations for multi-entropy, in order to determine whether similar exponents appear in $\alpha_\kappa$ and to perform a detailed finite-size scaling analysis, is an important direction for future work.

\begin{figure}[tbp]
    \centering
    \includegraphics[width=0.45\textwidth]{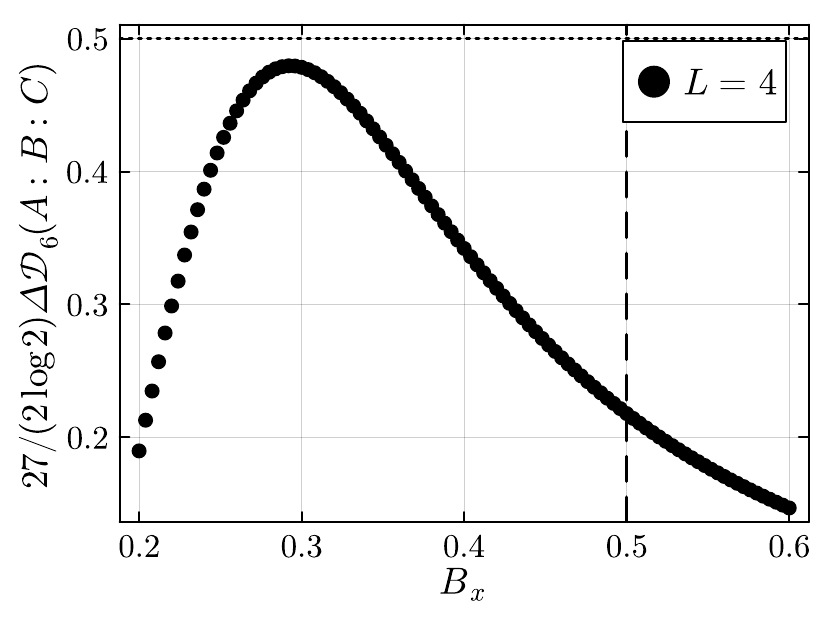}
    \caption{
        The rescaled dihedral excess $27/(2\log{2})\Delta\mathcal{D}_6$ for the adjacent setup with $\ell_A = \ell_B = 1$ and $L=4$. The vertical dashed line indicates the theoretical critical point $h_c=1/2$.
        The horizontal dotted line indicates the CFT prediction $c=1/2$.
    }
    \label{fig:Ising_adjacent_dihedral}
\end{figure}

The excess of dihedral measure is also computed for $L = 4$, as shown in Fig.~\ref{fig:Ising_adjacent_dihedral}.
Despite the small system size, the peak value is only 4\% below the CFT prediction~\eqref{diheu} with $n=3$,
\begin{equation}
    \Delta\mathcal{D}_6 = \frac{2c}{27}\log{2}.
\end{equation}

\begin{table}
    \centering
    \caption{
        Numerical and theoretical values of the critical point $h_c$, the adjacent multi-entropy excess $\kappa_c$, and the central charge $c$.
        Numbers in parentheses for numerical values represent one standard deviation.
    }
    \label{tab:adjacent_scaling}
    \begin{ruledtabular}
        \begin{tabular}{lrrr}
            & $h_c$ & $\kappa_c$ & $c$ \\
            Numerical & $0.5055(62)$ & $0.0871(26)$ & $0.5025(148)$ \\
            Theoretical & $0.5$ & $(c/4)\log{2} \simeq 0.0866$ & $0.5$ \\
        \end{tabular}
    \end{ruledtabular}
\end{table}

\subsubsection{Disjoint setup}
We now turn to the disjoint setup and analyze the dependence of entropic quantities on the separation $\ell$ between two subsystems $A$ and $B$.

\begin{figure*}[tbp]
    \centering
    \includegraphics[width=0.45\textwidth]{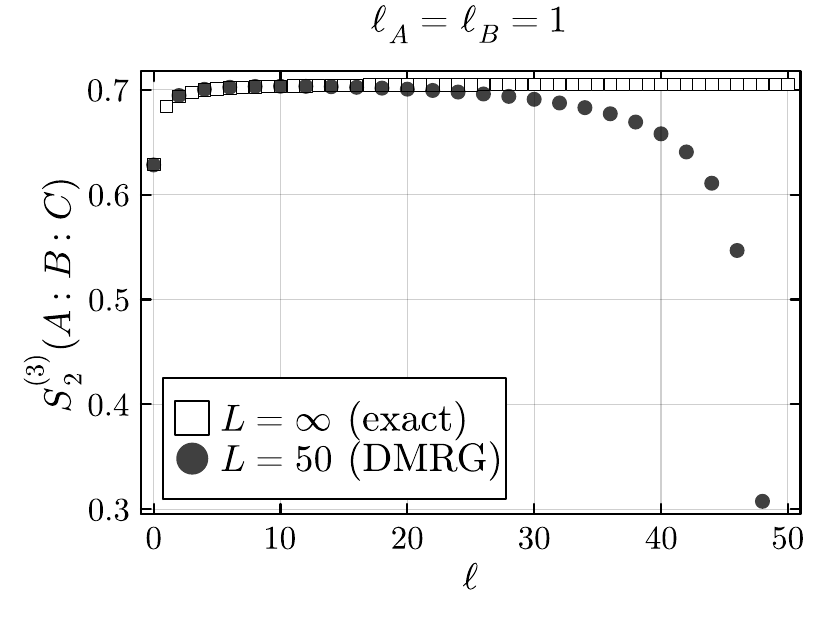}
    \includegraphics[width=0.45\textwidth]{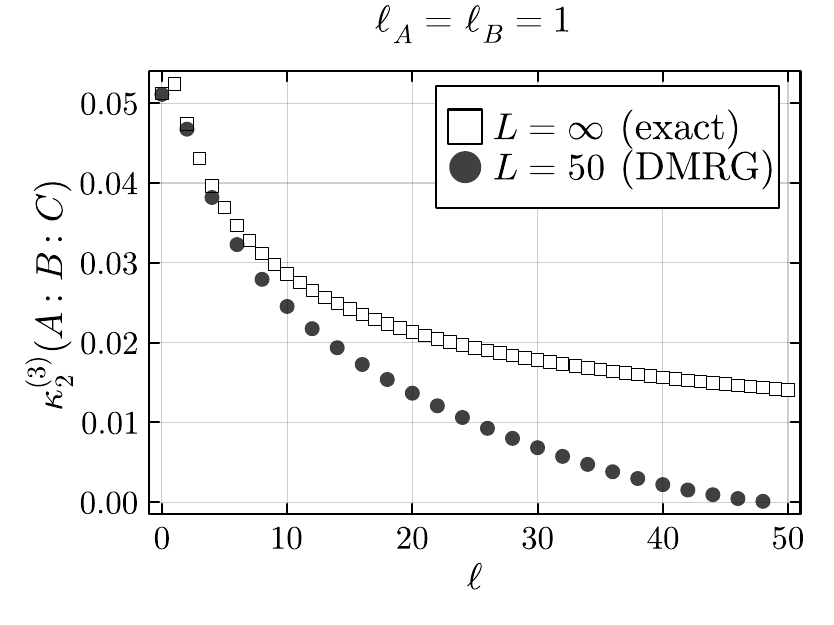}
    \includegraphics[width=0.45\textwidth]{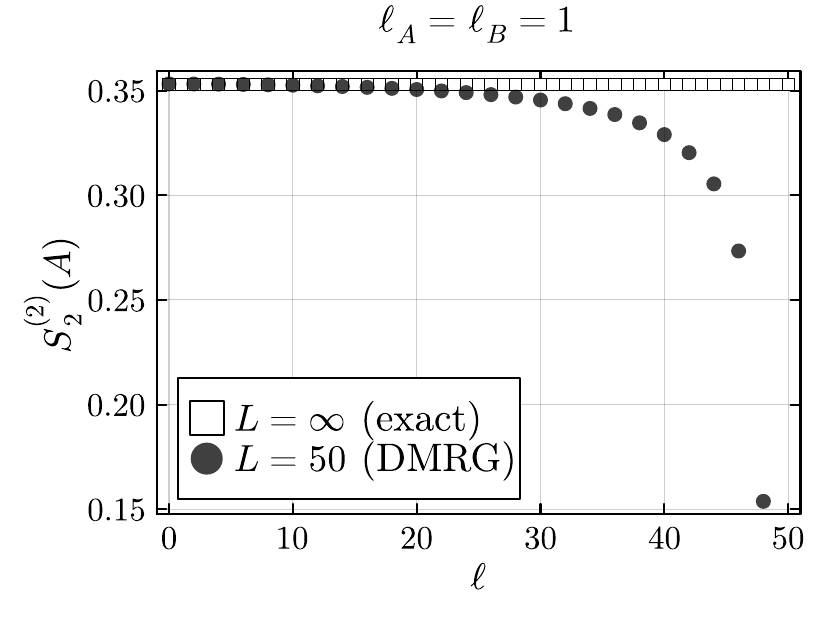}
    \includegraphics[width=0.45\textwidth]{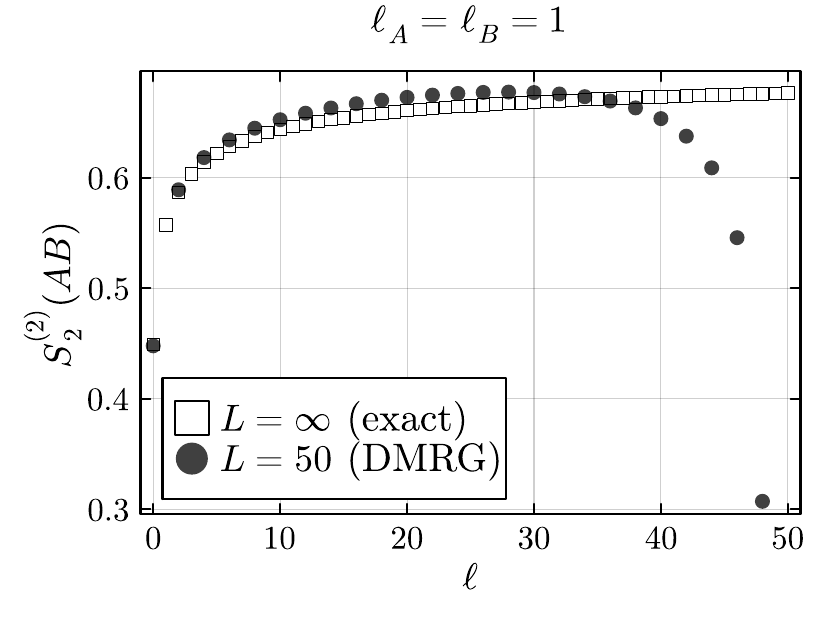}
    \caption{
        Dependence of entropic quantities on the subsystem separation $\ell$ in the disconnected setup with $\ell_A = \ell_B = 1$.
        White squares represent exact analytical results for the infinite system ($L \to \infty$) with integer values of $\ell$.
        Black circles indicate tensor network results for a finite open chain ($L = 50$) with even values of $\ell$.
        The transverse field is fixed at the theoretical critical point $h_c = 1/2$.
        \textit{Top left:}~multi-entropy $S_2^{(3)}$.
        \textit{Top right:}~multi-entropy excess $\kappa_2^{(3)}$.
        \textit{Bottom left:}~single-site Rényi entropy $S_2^{(2)}(A)$.
        \textit{Bottom right:}~two-site Rényi entropy $S_2^{(2)}(AB)$.
    }
    \label{fig:Ising_disconnected_rdep_lattice}
\end{figure*}

\paragraph{Benchmark against exact two-site results.}
Let us first consider the case of $\ell_A = \ell_B = 1$ at the true critical point $h_c = 1/2$, which allows us to compare the numerical results with exact analytical results derived in \ref{sec:exact}.
Fig.~\ref{fig:Ising_disconnected_rdep_lattice} compares the exact analytical results in the thermodynamic limit ($L \to \infty$) with numerical results obtained from the tensor network calculation on a finite open chain ($L=50$).
These highlight the accuracy of tensor network calculations in capturing multi-partite entanglement at short subsystem separations and deviations arising from boundary effects as the subsystems approach the boundaries.

In the top left panel of Fig.~\ref{fig:Ising_disconnected_rdep_lattice}, we present the multi-entropy $S_2^{(3)}$, which exhibits saturation to a constant value \eqref{eq:exact_distantlimit} as $\ell$ becomes large in the exact result.
The numerical data of multi-entropy shows good agreement with the exact result for small separation $\ell \lesssim 10$, but deviates for larger $\ell$ due to suppression near boundaries.
The top right panel shows the multi-entropy excess $\kappa_2^{(3)}$.
While the numerical results agree with the exact result for small separations ($\ell = 0, 2$), they exhibit a faster decay as the subsystems approach the boundaries.
The bottom left panel shows the single-site Rényi entropy $S_2^{(2)}(A)$, which is independent of $\ell$ in the infinite system.
The numerical result for the single-site Rényi entropy agrees with the exact result for $\ell \lesssim 10$.
The bottom right panel shows the two-site Rényi entropy $S_2^{(2)}(AB)$, which asymptotically approaches $2S_2^{(2)}(A)$ for exact results.
The numerical values match the exact result for $\ell \lesssim 4$.

\begin{figure*}[tbp]
    \centering
    \includegraphics[width=0.45\textwidth]{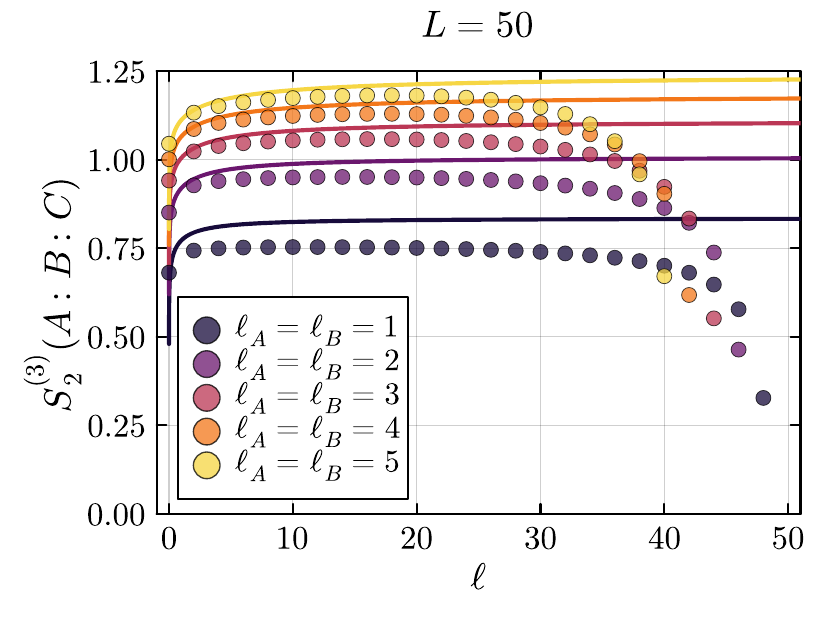}
    \includegraphics[width=0.45\textwidth]{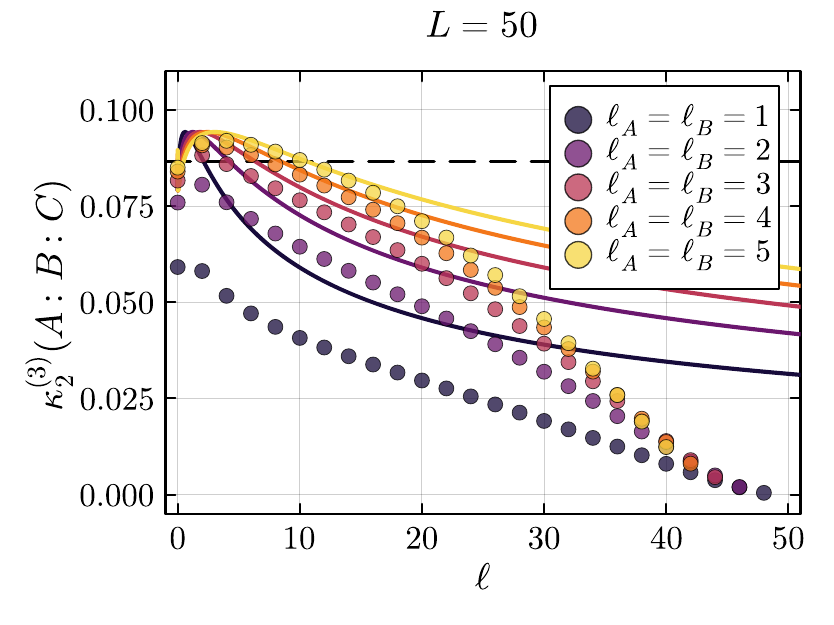}
    \includegraphics[width=0.45\textwidth]{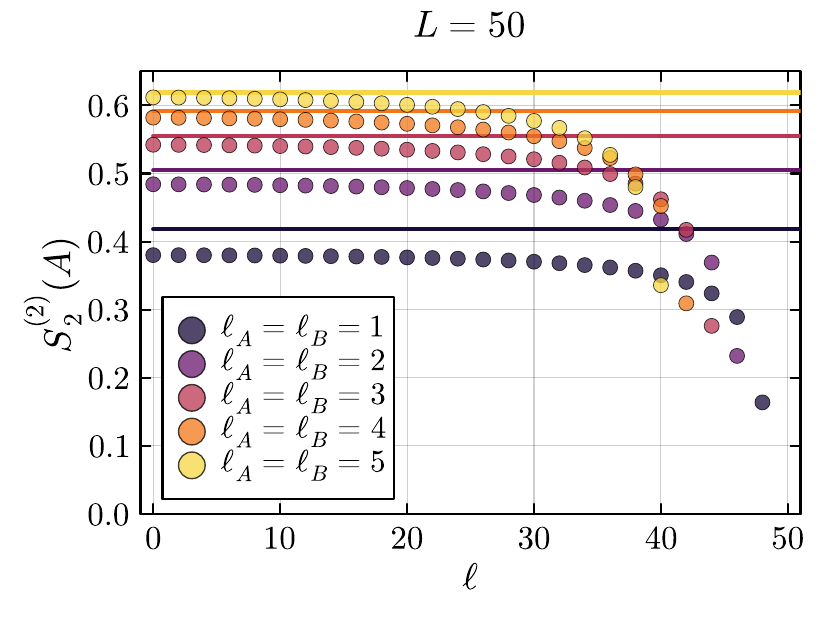}
    \includegraphics[width=0.45\textwidth]{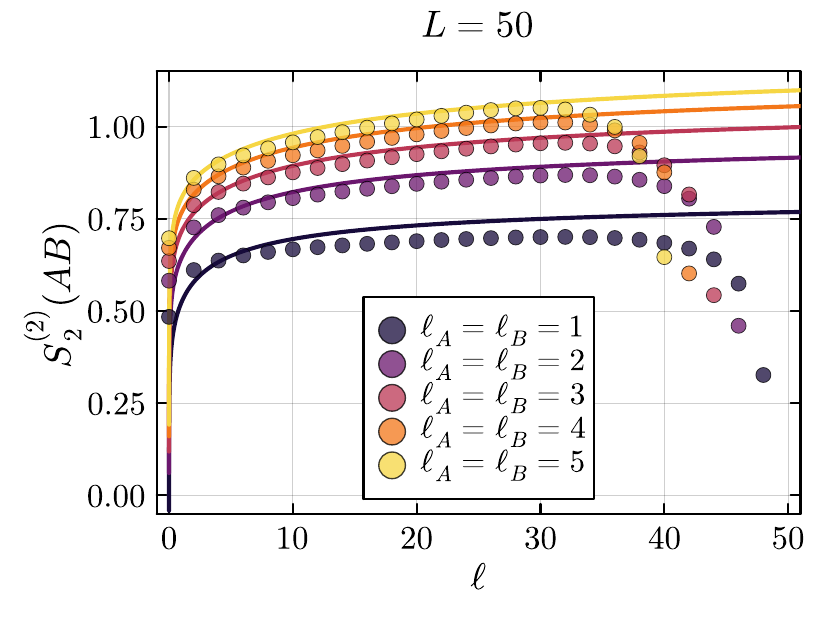}
    \caption{
        Dependence of entropic quantities on the subsystem separation $\ell$ in the disconnected setup with $\ell_A = \ell_B = 1, 2, 3, 4, 5$.
        Circles indicate tensor network results for a finite open chain ($L = 50$) with even values of $\ell$.
        Lines represent the corresponding CFT prediction for the infinite system ($L \to \infty$) with the cutoff-dependent shift~\eqref{eq:shift_value}.
        The transverse field is fixed at the value that maximizes the adjacent multi-entropy excess $\kappa_2^{(3)}$ for $\ell_A = \ell_B = 5$, which is $h \approx 0.483$.
        \textit{Top left:}~multi-entropy $S_2^{(3)}$. 
        \textit{Top right:}~multi-entropy excess $\kappa_2^{(3)}$. 
        \textit{Bottom left:}~single-interval Rényi entropy $S_2^{(2)}(A)$.
        \textit{Bottom right:}~two-interval Rényi entropy $S_2^{(2)}(AB)$.
    }
    \label{fig:Ising_disconnected_rdep}
\end{figure*}

\paragraph{Larger blocks and comparison with CFT.}
Fig.~\ref{fig:Ising_disconnected_rdep} extends the analysis to symmetrically enlarged blocks with $\ell_A = \ell_B = 1, 2, 3, 4, 5$ on a finite open chain ($L=50$) with even values of $\ell$.
The transverse field is fixed to $h \approx 0.483$, the value that maximizes the adjacent multi-entropy excess $\kappa_2^{(3)}$ for the largest calculated subsystem size $\ell_A = \ell_B = 5$.
The cutoff-dependent shift for the CFT prediction is determined as described in~\ref{ssec:fit_disjoint}.

\begin{figure}[tbp]
    \centering
    \includegraphics[width=0.45\textwidth]{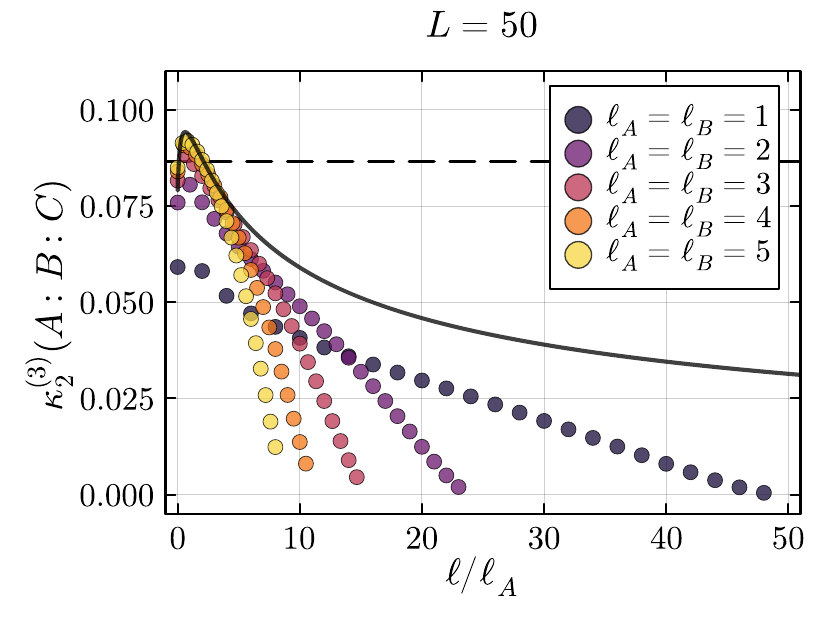}
    \caption{
        Multi-entropy excess $\kappa_2^{(3)}$ for the disconnected setup with $\ell_A = \ell_B = 1, 2, 3, 4, 5$ as a function of the ratio $\ell/\ell_A$.
        The transverse field is fixed at the value that maximizes the adjacent multi-entropy excess $\kappa_2^{(3)}$ for $\ell_A = \ell_B = 5$, which is $h \approx 0.483$.
        Circles indicate tensor network results for a finite open chain ($L = 50$) with even values of $\ell$.
        The black line represents the CFT prediction~\eqref{IsingDS} for the infinite system ($L \to \infty$).
    }
    \label{fig:Ising_disconnected_rdep_kappa_rescaled}
\end{figure}

As shown in Fig.~\ref{fig:Ising_disconnected_rdep}, the numerical results agree with the analytical predictions at small separations $\ell$, with deviations appearing at larger separations due to boundary effects.
Notably, the agreement with the CFT improves systematically as the subsystem size $\ell_A = \ell_B$ incleases.
Plotting the multi-entropy excess as a function of the scaled ratio $\ell/\ell_A$ as shown in Fig.~\ref{fig:Ising_disconnected_rdep_kappa_rescaled}, a universal curve emerges in the regime where $\ell/\ell_A$ is small and $\ell_A$ is large, consistent with the CFT prediction~\eqref{IsingDS}.
The ratio $\ell/\ell_A$ introduced here is related to the cross-ratio $\eta$ as
\begin{equation}
    \eta = \frac{1}{(1 + \ell / \ell_A)^2}.
\end{equation}

\section{Conclusions}
In this article, we have computed the multi-entropy and dihedral measures, which are tractable measures for multi-partite entanglement. They are constructed out of multi-invariants and can be regarded as generalizations of Rényi entanglement entropy by extending the group symmetries of replicas. The main advantage of these quantities is that they probe multi-partite quantum corrections which cannot be computed from the conventional quantum information measures like von-Neumann and Rényi entropy. They are labeled by the Rényi index (or replica number) $n$ as in the Rényi entropy.

We first presented basic definitions and properties of the multi-entropy and dihedral measure. In particular, we focused on the tripartite case $\tq=3$, which is defined by decomposing the total system into three subsystems $A$, $B$, and $C$. We introduced the excess of multi-entropy $\kappa^{(3)}_n$ and that of the dihedral measure $\Delta{\cal D}_{2n}$ by subtracting the bi-partite entanglement. We computed them explicitly for GHZ, W, and Werner states and found that they show characteristic behaviors.

After this, we considered computations of the multi-entropy and dihedral measure at quantum critical points. We performed both field-theoretic calculations in 2d CFTs and the numerical calculations in the free scalar and transverse-field Ising model on a one-dimensional lattice.
When the subsystem $A$ and $B$ are adjacent, the CFT analysis shows that the excesses $\kappa^{(3)}_2$ and $\Delta{\cal D}_{2n}$ take the universal form (\ref{generalkab}) and (\ref{diheu}). The numerical results of the Ising model successfully reproduced these results.

However, the field theory analysis tells us that the free scalar CFT is an exception of  (\ref{generalkab}) in that we have $\kappa^{(3)}_2=0$, which is due to the zero mode of the scalar field, peculiar to the non-compact CFT. Indeed, we confirmed this numerically in our lattice model calculations. In addition, we find 
$\kappa^{(3)}_3\simeq -0.14$ and $\kappa^{(3)}_4\simeq -0.2$ in the free scalar CFT.
For the dihedral measure, the numerical result also shows $\Delta{\cal D}_{4}=0$, while $\Delta{\cal D}_{6}\simeq 0.028$ 
and $\Delta{\cal D}_{8}\simeq 0.031$ in this CFT.

We also studied the multi-entropy when the subsystems $A$ and $B$ are separated by a distance $d$. We evaluated this analytically by the conformal field theoretic calculations for the free Majorana fermion (Ising) CFT, free scalar CFT, and the holographic CFTs, which qualitatively differ from each other. In the free scalar field case, we find that the excess $\kappa^{(3)}_2$ is independent of $d$ and is always vanishing. In the Majorana fermion CFT, it shows the power law decay $\kappa^{(3)}_2\propto\frac{1}{\s{d}}$, while it decays much more quickly in the holographic CFTs. This implies that the multi-entropy provides a useful quantity which can distinguish different quantum critical points. Our numerical calculations in the free scalar field theory on a lattice does not only perfectly reproduce the above field theoretic result $\kappa^{(3)}_2=0$ for any $d$ but also shows that $\kappa^{(3)}_3$ and $\kappa^{(3)}_4$ are negative valued functions of $d$. We also numerically computed  $\kappa^{(3)}_2$ as a function of $d$ in the Ising model at the critical point via a tensor network method. We tested the validity of our calculation by comparing it with the exact results when each of $A$ and $B$ consists of a single site. Our numerical result agrees with that of the $c=\frac{1}{2}$ Majorana fermion CFT for small values of $d$. For large $d$, our result deviates from the CFT result due to the boundary effect, as expected. In conclusion, our present calculations demonstrate that the multi-entropy and dihedral measure are new useful probes to characterize quantum critical points from the viewpoint of multi-partite entanglement.

There are several interesting future directions. It would be important to better understand what quantum correlations the multi-entropy, dihedral measures, and more general measures obtained from the multi-invariants, precisely estimate in terms of quantum information theory. It is also intriguing to explore how these quantities can be computed in holography. Since in order for the measures to be related to pure $AdS$, i.e. free of conical singularities, it is necessary to consider something analogous to the $n\rightarrow 1$ limit for holographic entanglement entropy in the case of general multi-invariants.   As such it would be desirable to develop both the numerical interpolations of the $n=1$ limit and analytical calculations in field theories for wider values of $n$. At the same time, it would also be interesting to extend our analysis for the basic quantum many-body systems to various other models of quantum critical points, in particular, those in strongly correlated systems.

\vspace{5mm}
{\it Acknowledgements:}
We are grateful to Takanao Ishii, Shinsei Ryu, Haruki Shimizu, Tokuro Shimokawa, and Shinji Takeda for valuable discussions.
This work is supported by MEXT KAKENHI Grant-in-Aid for Transformative Research Areas (A) through the ``Extreme Universe'' collaboration: Grant Number 21H05187. TT is also supported by JSPS Grant-in-Aid for Scientific Research (B) No.~25K01000.
KT is supported by Grant-in-Aid for JSPS Fellows No.~25KJ1455.
Part of the numerical computations in this work were performed on Yukawa-21 at Yukawa Institute for Theoretical Physics, Kyoto University.

\appendix
\section{Theta function identities}

We list useful identities of theta functions which were employed in section \ref{sec:multi-dis}:
\ba
&&\eta(2\tau)=2^{-\frac{2}{3}}\left(\theta_2(\tau)\right)^{\frac{2}{3}}\left(\theta_3(\tau)\right)^{\frac{1}{6}}\left(\theta_4(\tau)\right)^{\frac{1}{6}},\no
&&\eta(\tau)=2^{-\frac{1}{3}}\left(\theta_2(\tau)\right)^{\frac{1}{3}}\left(\theta_3(\tau)\right)^{\frac{1}{3}}\left(\theta_4(\tau)\right)^{\frac{1}{3}},\no
&& \theta_2(2\tau)=\s{\frac{\left(\theta_3(\tau)\right)^2
-\left(\theta_4(\tau)\right)^2}{2}},\no
&&\theta_3(2\tau)=\s{\frac{\left(\theta_3(\tau)\right)^2
+\left(\theta_4(\tau)\right)^2}{2}},\no
&&\theta_4(2\tau)=\s{\theta_3(\tau)\theta_4(\tau)}. \label{thetai}
\ea

\section{Numerical fit for transverse field Ising model}
\subsection{Adjacent setup}\label{ssec:fit_adjacent}
To precisely extract the pseudo-critical point $h_{\text{pc}}(L)$ and the peak value $\kappa_{\text{pc}}(L)$ from adjacent multi-entropy excess, we fit the numerical data within a window centered on the numerical peak.
We select a window of 19 data points (width $\varDelta h = 0.018$) for all system sizes.
We fit the data point inside the window using a quadratic form
\begin{equation}
    \kappa(L, h) \simeq \kappa_{\text{pc}}(L) - \frac{a(L)}{2} [h - h_{\text{pc}}(L)]^2.
\end{equation}

To extrapolate the true critical point $h_c$ and the limiting value $\kappa_c$ from the pseudo critical point $h_{\text{pc}}(L)$ and the peak value $\kappa_{\text{pc}}(L)$, we fit the data for $L \geq 8$ using the form
\begin{subequations}
\begin{align}
    h_{\text{pc}}(L) = h_c - b_h L^{-\alpha_h}, \\ \kappa_{\text{pc}}(L) = \kappa_c - b_\kappa L^{-\alpha_\kappa}.
\end{align}
\end{subequations}
The results are summarized in Table~\ref{tab:adjacent_scaling} in the main text.

\subsection{Disjoint setup}\label{ssec:fit_disjoint}

Colored lines in Fig.~\ref{fig:Ising_disconnected_rdep} represent the corresponding CFT prediction for the infinite system ($L \to \infty$), where the cutoff-dependent shift is applied consistently across the multi-entropy and Rényi entropies, while no adjustment is made for multi-entropy excess since it is independent of the cutoff.
The value of the shift is determined based on the relation
\begin{equation}
    S_2^{(2)}(A) \vert_{\text{lattice}} - \frac{c}{4}\log{\ell_A} = - \frac{c}{4} \log{\epsilon}.\label{eq:shift}
\end{equation}
By fitting the left-hand side of \eqref{eq:shift} at $\ell = 0$ to the form $S_\infty - b_S \ell_A^{-\alpha}$, we find the intercept in the $\ell_A\to \infty$ limit.
The intercept determines the appropriate value for the shift of the CFT prediction for single-interval entropies as
\begin{equation}
    -\frac{c}{4} \log{\epsilon} = 0.4196(3) \label{eq:shift_value}
\end{equation}
The shift for two-interval entropies is twice the value for single-interval entropies obtained here.

\bibliography{main.bib}


\end{document}